\documentclass{aa}

\usepackage{natbib}
\bibpunct{(}{)}{;}{a}{}{,} 
\usepackage{color}
\usepackage{graphicx}
\usepackage{amssymb}
\usepackage{amsmath}

\begin{document}

\renewcommand{\vec}[1]{\bmath{#1}}
\newcommand{\sign}{\textrm{sign}}
\newcommand{\pd}[2]{\frac{\partial #1}{\partial #2}}
\newcommand{\DS}{\displaystyle}
\newcommand{\HALF}{\frac{1}{2}}
\newcommand{\mathi}{\rm i}

\renewcommand{\vec}[1]{\mathbf{#1}}
\newcommand{\hvec}[1]{\hat{\mathbf{#1}}}
\newcommand{\av}[1]{\left<#1\right>}
\newcommand{\red}[1]{\color{red} #1 \color{black}}
\newcommand{\blue}[1]{\color{blue} #1 \color{black}}

\title{3D Relativistic MHD numerical simulations of X-shaped radio sources}

\author
       { P. Rossi\inst{\ref{inst1}}\thanks{E-mail:rossi@oato.inaf.it} \and G. Bodo\inst{\ref{inst1}}  \and A. Capetti\inst{\ref{inst1}} \and  S. Massaglia\inst{\ref{inst2}} }
\institute{INAF/Osservatorio Astrofisico di Torino, Strada Osservatorio 20, 10025 Pino Torinese, Italy \label{inst1}
\and Dipartimento di Fisica  Universit\`a degli Studi di Torino, Via Pietro Giuria 1, 10125 Torino, Italy  \label{inst2}
}

\authorrunning{P. Rossi et al.}

\date{Accepted ??. Received ??; in original form ??}


\titlerunning{3D RMHD X-shaped radio sources}

\label{firstpage}

\abstract
{A significant fraction of extended radio sources presents a peculiar X-shaped radio morphology: in addition to the classical double lobed structure, radio emission is also observed along a second axis of simmetry in the form of diffuse wings or tails. In a previous investigation  we showed the existence of a connection between the radio morphology and the properties of the host galaxies.  Motivated by this connection we  performed two-dimensional numerical simulations showing that X-shaped radio sources may naturally form as a jet propagates along the major axis a highly elliptical density distribution, because of the fast expansion of the cocoon along the minor axis of the distribution.
}{
We intend to extend our analysis by performing three-dimensional numerical simulations and investigating the role of different parameters is determining the formation of the X-shaped morphology.
}{
The problem is addressed by numerical means, carrying out three-dimensional relativistic magnetohydrodynamic simulations of  bidirectional jets propagating in a triaxial density distribution. 
}{
We show that  only jets with power $\lesssim 10^{44}$ erg s$^{-1}$ can give origin to an X-shaped morphology and that a misalignment of $30^o$ between the jet axis and the major axis of the density distribution is still favourable to the formation  of this kind of morphology. In addition we compute synthetic radio emission maps and polarization maps.}
{
In our scenario for the formation of X-shaped radio sources only low power FRII can give origin to such kind of morphology. Our synthetic emission maps show that the different observed morphologies of X-shaped sources can be the result of similar structures viewed under different perspectives.
}

\maketitle

\section{Introduction}
%
%
%
 \citet{Fanaroff1974} classified the extended radio sources on the basis of their morphologies and identified two classes, since then called Fanaroff-Riley I and II. The characteristic structure of FRII source is dominated by two hot spots located at the edges of the radio lobes that, in most cases, show bridges of emission linking the core to the hot spots. A significant fraction of FRII, about 10\%, show a strong distortion in the bridges, that can be mirror symmetric (or C-shaped) when the bridges bend away from the galaxy in the same direction, or center symmetric, when they bend in opposite direction forming an X-shaped. In many X-shaped sources the radio emission along the secondary axis, although more diffuse, is still quite well collimated and can be even more  extended than the main lobed structure. 
 In the past 40 years, many authors suggested different interpretations of X-shaped radio galaxies,
\citet{Ekers1978} suggested that the tails of radio emission in one of these sources, NGC 326, are the result of the trail caused by a secular jet precession \citep[see also][]{Rees1978}. A similar model accounts for the morphology of 4C 32.25 \citep{Klein1995}. In a similar line, \citet{Wirth1982} noted that a change in the jet direction can be caused by gravitational interaction with a companion galaxy. \citet{Dennett-Thorpe2002}, from the analysis of spectral variations along the lobes, proposed that the jet reorientation occurs over short time scales, a few Myr, and are possibly associated to instabilities in the accretion disk that cause a rapid change in the jet axis. In all these models, the secondary axis of radio emission represents a relic of the past activity of the radio source. An alternative interpretation was suggested by \citet{Leahy1984} and \citet{Worrall1995}. They emphasize the role of the external medium in shaping radio sources, suggesting that buoyancy forces can bend the back-flowing material away from the jet axis into the direction of decreasing external gas pressure. 
\citet{Capetti2002} (Paper I)  proposed a different scenario based on evidence for a strong connection between the properties of the radio emission and those of the host galaxy, a comparison that has been overlooked in the past, but that provides crucial new insights on the origin of X-shaped radio-sources. A more complete summary of the models proposed for the origin of X-shaped radiogalaxies can be found in \citet{Gopal12}. 
In Paper I we compared the radio and host galaxy orientation of X-shaped radio sources showing that there is a close alignment between the radio wings and the galaxy minor axis. We also showed that X-shaped sources occur exclusively in galaxies of high ellipticity.  These results have been confirmed by the analysis of a larger sample by \citet{Gillone16}.
They selected, from an initial list of 100 X-shaped RGs from \citet{Cheung07}, the 53 galaxies with the better defined wings and with available SDSS images; in 22 sources it was possible to measure the optical position angle. The orientation of the secondary radio structures shows a strong connection with the optical axis, with all (but one) wing forming an angle larger than 40$^o$ with the host major axis.
 Motivated by the observational evidences of the relation between the radio morphology and the optical galactic structure, in Paper I we presented the results of 2D numerical simulations of the propagation of a jet in a stratified medium with an elliptical symmetry.   The simulations show that the radio source evolution and morphology is strongly influenced by the distribution of the external density and by the orientation of the jet propagation with respect to this distribution. If the jet propagates along the major axis of the density distribution, initially it inflates an almost spherical cocoon that becomes soon elongated in the jet direction, but also substantially expands along the minor axis  due to the faster decrease of the external density in this direction. In this way we can observe the formation of a typical X-shape morphology.  In Paper I we showed that, on the contrary, when the jet propagates along the minor axis we don't observe the wing formation and we have a classical double radio source. 

 A limitation of the investigation performed in Paper I is the dimensionality.
More recently \citet{Hodges-Kluck2011} performed three-dimensional simulations and investigated the effect of a 3D elliptical stratification on the propagation of jets by using Newtonian hydrodynamical  simulations. They emphasized the importance of ellipticity of the host galaxy and they stressed how the hot atmosphere may play a dominant role in shaping the morphology of radiogalaxies. 
In this paper, we intend to reconsider our previous work proposing a much more detailed numerical approach and extend  the results obtained by \citet{Hodges-Kluck2011} by including magnetic and relativistic effects,  assuming a triaxial equilibrium density distribution of the  external medium.  Our approach differs from that of \citet{Hodges-Kluck2011}  also because they consider the long term evolution of the radio sources looking at the effect of switching-off the jet and of buoyancy forces, while we study the  first part of the evolution and consequently also our spatial scales are smaller. Our purpose is to investigate how the formation of the wings depends on the jet parameters like their Lorentz factor $\gamma_j$ and their density ratio $\nu$ with respect to the external medium.  Moreover we want also to explore  the consequence of misalignment between jet axis and galaxy major axis. 

In the next section we will describe the numerical setup and the set of simulations that we performed, in  Section 3 we present our results focusing in particular on constraining the jet parameters for the formation of the wings and finally our conclusions are outlined in the last section.

\section{Problem Description}
%
%
%

Numerical simulations are carried out by solving the equations of relativistic magnetohydrodynamic
\begin{eqnarray}
\partial_\mu \left( \rho u_\mu \right) = 0,  \\ 
\partial_\mu \left(  w u^\mu u^\nu - b^\mu b^\nu + p \eta^{\mu\nu}  \right) = 0,  \\
\partial_\mu \left(  u^\mu b^\nu  - u^\nu b^\mu \right) = 0,
\end{eqnarray}
where $\rho$ is the rest mass density, $u^\mu \equiv \gamma (1, \vec{v}) $ and $b^\mu = (b^0, \vec{B}/\gamma + b^0 \vec{v} )$ are respectively the four-velocity and covariant magnetic field written in terms of the three-velocity $\vec{v}$ and laboratory magnetic field $\vec{B}$. A flat metric $\eta^{\mu \nu} = \rm{diag}(-1, 1, 1, 1)$ is considered. In the previous equations $w = w_g + b^2$ and $p = p_g + b^2/2$ express the total enthalpy and total pressure in terms of their thermal ($w_g$ and $p_g$) and magnetic field contributions ($b^2 = b^\mu b_\mu$), respectively.  We assume a single-species relativistic perfect fluid (the Synge gas) described by the approximated equation of state proposed by \citet{Mignone05}.

\begin{figure*}[t!]
   \centering
   \includegraphics[width=12cm]{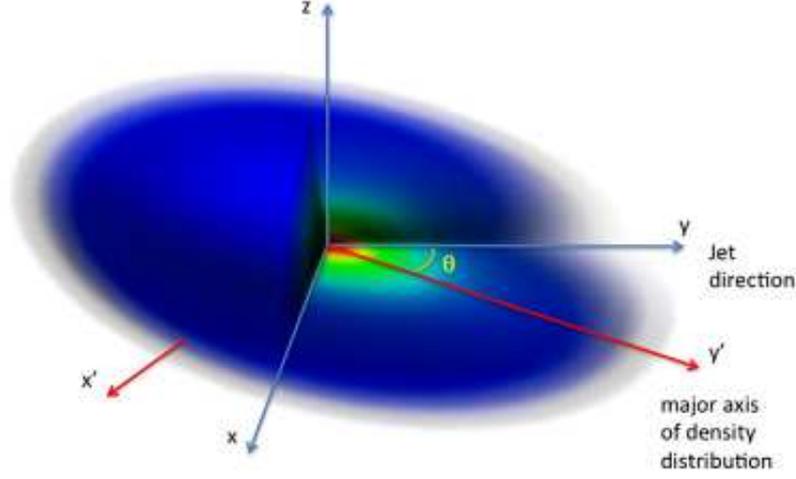} 
\caption{Sketch of the simulation setup showing a representation of the initial density distribution and the relative orientation of the jets.
 }
   \label{fig:setup}
\end{figure*}

\begin{figure}[h!]
   \centering
   \includegraphics[width=8cm]{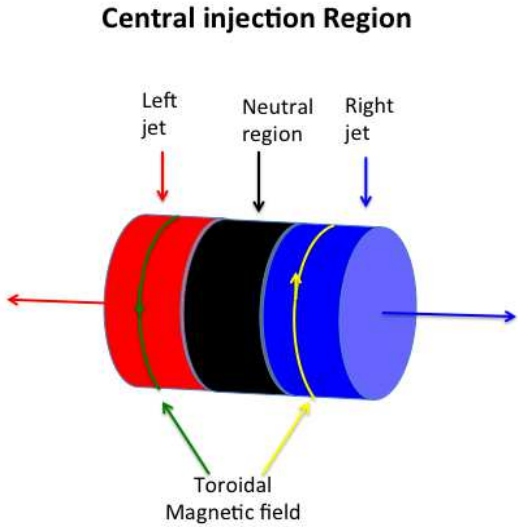} 
\caption{ Sketch of the cylindrical injection region which represents an internal boundary in which the variables are kept fixed to the injection values. In the black part velocity and magnetic field are kept equal to zero, in the blue and red regions velocities are opposite and from these two regions two opposite jets are injected.  The blue and red arrows show the velocity directions, while the arrows on the yellow and green lines show the direction of the toroidal magnetic field. }
   \label{fig:internalboundary}
\end{figure}

We study  the propagation of bidirectional jets flowing into a triaxial ellipsoidal atmosphere initially isothermal and in hydrostatic equilibrium. The simulations are performed on a three-dimensional cartesian grid with the $y$ axis aligned with the jet direction. The computational domain is initially filled with a  density distribution chosen as a $\beta-$model \citep{Cavaliere1976}  deformed into a triaxial shape by taking  different effective core radii along the three axes. The effective core radii $a_x$, $a_y$, $a_z$ and all the lengths are measured in unit of a fictitious core radius $r_c$, whose value can be chosen to be consistent with observational data. The initial density distribution  is given by
\begin{equation}
\rho = \frac{\rho_0}{[1 + (x'/a_x)^2 + (y'/a_y)^2 + (z/a_z)^2]^{3 /2 \beta}}
\label{eq:density}
\end{equation}
where $\rho_0$ is the central density, 
\begin{equation}
x'  = x \cos(\theta) - y \sin(\theta)
\end{equation}
\begin{equation}
y'  = x \sin(\theta) +  y \cos(\theta)
\end{equation}
are the coordinate along the axes of the ellipsoidal distribution and  $\theta$ is the angle between the jet and the major axis of the atmosphere. A sketch of this setup is shown in Fig. \ref{fig:setup}.   In all the simulations we used   $a_x = 1/3$, $a_y = 2/3$ and $a_z = 1/4$. The pressure distribution follows the density distribution and equilibrium is maintained by an external gravitational potential. Two jets, with initial radius $ r_j = 0.05 r_c$,  flowing in opposite directions  are injected at the center of the domain.   No magnetic field is present in the initial configuration at $t = 0$, and a toroidal  magnetic field is injected along with the jets. The injection region is a cylinder of radius $r_j$ and length $0.04 r_c$   aligned with the $y$ axis in which the variables are not evolved but they are kept fixed at the injection values.  As shown in Fig. \ref{fig:internalboundary} this region is divided in three part, a central part (shown in black in the figure) in which velocity and magnetic field are zero, and two side regions (shown in red and blue in the figure) in which the jet velocities are opposite and the magnetic field is defined as
\begin{equation}
B_x = -B_t  r \sin(\phi)
\end{equation}
\begin{equation}
B_y = B_t r \cos{\phi}
\end{equation}
in the blue region and is reversed in the red region.  Here $r$ and $\phi$ are polar coordinates in a plane normal to the $y$ axis and the maximum field strength $B_t$ is fixed by the value of the jet magnetization, $\sigma$, which represents the ratio between the Poynting flux and the matter energy flux, in the jets.  A jet with a purely toroidal field is potentially unstable to pinch and kink magnetohydrodynamic instabilities, however the effect of these instabilities starts to show up only at late times where some jet wiggling can be observed.

The two jets are slightly perturbed at their base, in order to destroy all symmetries, the following evolution is independent from the details of the perturbation. The  parameters that characterize this configuration  are the jet Lorentz factor $\gamma_j$, the density ratio $\nu$,  between the jet density  and the  central density,  the jet temperature and the magnetization $\sigma$,    The temperature of the external medium is fixed by the jet Mach number $M_j$, which is 400 for all simulations.  In Table \ref{parset} we show all the parameters for the simulations that we present in this paper.    In the sixth column of  Table \ref{parset} we give the values of the jet energy flux defined by
\begin{equation}
L_j = \pi r_j^2 \rho_j \gamma_j^2 c^3  = 2.5 \times 10^{-3} \pi \rho_0 r^2_c \nu \gamma_j^2 c^3 
\label{eq:lj}
\end{equation}
and computed for the representative values $r_c = 4  {\rm kpc}$ and $\rho_0 = 1  \hbox{cm}^{-3}$, which are consistent with observational data.  
The details of the injection setup have of course some influence on the characteristics of jet propagation. However, we can expect limited differences on the large scale structure, also in view of the fact that the perturbations imposed at the base will quite likely cause the jet to lose memory of the details of the initial state, while propagating into the ambient medium.

\begin{table*}  
\centering
\begin{tabular}{ccrccccc}\hline                                        
Case  &  $\theta$ & $\gamma_j$ & $\nu$ & $\sigma$  & \multicolumn{1}{c}{$L_j $} & Domain & Grid \\ 
\hline\hline\noalign{\medskip}
A        &       0   &      $5$           &     $1 \times 10^{-4} $                      &   $10^{-3}$      &   $1.6 \times   10^{46} $ &      $2\times4\times2$ & $384\times768\times384$ \\ 
B        &       0   &      $2.5$          & $4 \times 10^{-4}$            &    $10^{-3} $                &    $1.6 \times   10^{46} $  &      $2\times4\times2$ & $384\times768\times384$      \\ 
C        &      0   &      $1.25$       &  $2.66 \times 10^{-3}$     &    $ 10^{-3}  $                &   $1.6 \times   10^{46}  $   &     $4\times4\times4$ &  $768\times768\times768$ \\   
D        &     0   &       $5$            &   $ 1 \times 10^{-5} $                        &   $10^{-2}$     &   $ 1.6 \times  10^{45}  $  &     $4\times8\times4$  & $768\times1024\times768$   \\
E        &     0  &         $5$         &    $1 \times 10^{-6} $                      &    $10^{-2} $       &   $ 1.6 \times  10^{44} $   &      $4\times8\times4$   & $768\times1024\times768$   \\ 
F        &     $30^o$  &  $5$         &   $1 \times 10^{-6} $                         &   $10^{-2} $    & $  1.6 \times 10^{44}$     &       $4\times8\times4$& $768\times1024\times768$  \\
\noalign{\medskip}
\hline
\end{tabular}
\caption[]{Parameters  used in the simulation models. In the first column we have the case identifier, in the second column the angle between the jet axis and the major axis
of the triaxial density distribution, in the third column the jet Lorentz factor, in the fourth column the density ratio, in the fifth column the magnetization parameter $\sigma$, in the sixth
column the jet power, in the seventh column the size of the computational domain in units of $r_c$ and in the eighth column the grid size. \newline
$L_j$ is measured in erg s$^{-1}$ and is computed  for  
$r_c = 4$  kpc and $\rho_0 = 1$  cm$^{-3}$, the dependence of $L_j$ on $r_c$ and $\rho_0$ is given by Eq. \ref{eq:lj}}
\label{parset}
\end{table*}

The first three simulations (A, B and C) have constant energy fluxes, but different and decreasing Lorentz factors $\gamma_j$.  Cases A, D and E have fixed Lorentz factor $\gamma_j$, but decreasing density ratio and consequently decreasing energy flux. We have also to notice that cases D and E have a ten time higher magnetization than case A, the Poynting flux of case D is then equal to that of case A. Finally, cases $E$ and $F$ differ only for the angle $\theta$ between the jet direction and the major axis of the atmosphere.  In Table \ref{parset} we give, in the last two columns, for each case,  the size  of the computational domain and the grid size.  For all the simulations the grid is uniform in the central region ($1 \times 2 \times 1$) and is stretched in the outer parts in order to achieve larger distances. The simulations have been performed with the PLUTO code \citep{Mignone2007}, with a second order accurate linear reconstruction and HLL Riemann solver. 
 
\section{Results}
%
%

In the presentation of the results, all quantities are expressed in physical units, their dimensional values depend however on the choice of our unit of length $r_c$ and density $\rho_0$, which we have assumed $r_c = 4  \rm{kpc}$ and $\rho_0 = 1 \rm{cm}^{-3}$. A different choice for these units will give different values for all the relevant physical quantities, more precisely lengths scale as $r_c$, times scale also as $r_c$, pressure scales as $\rho_0$ and magnetic field strength scales as $\sqrt{\rho_0}$. 

In the first three simulations (A, B and C) we kept a constant jet power of $1.6 \times 10^{46}$ erg s$^{-1},$  but we  varied the Lorentz factor and the jet density, keeping $\gamma_j^2 \rho_j$, and hence the energy flux, constant. In all cases, at the injection point, the jet results to be overpressured with respect to the ambient medium, with increasing pressure going form case A to case C. In cases D and E, instead,  we kept a constant Lorentz factor but we decreased the jet density and therefore the jet power which results to be $1.6 \times 10^{45} $ erg s$^{-1}$ in case D and  $1.6 \times 10^{44} $ erg s$^{-1}$  in case E.
 
\begin{figure*}[t!]
   \centering
   \includegraphics{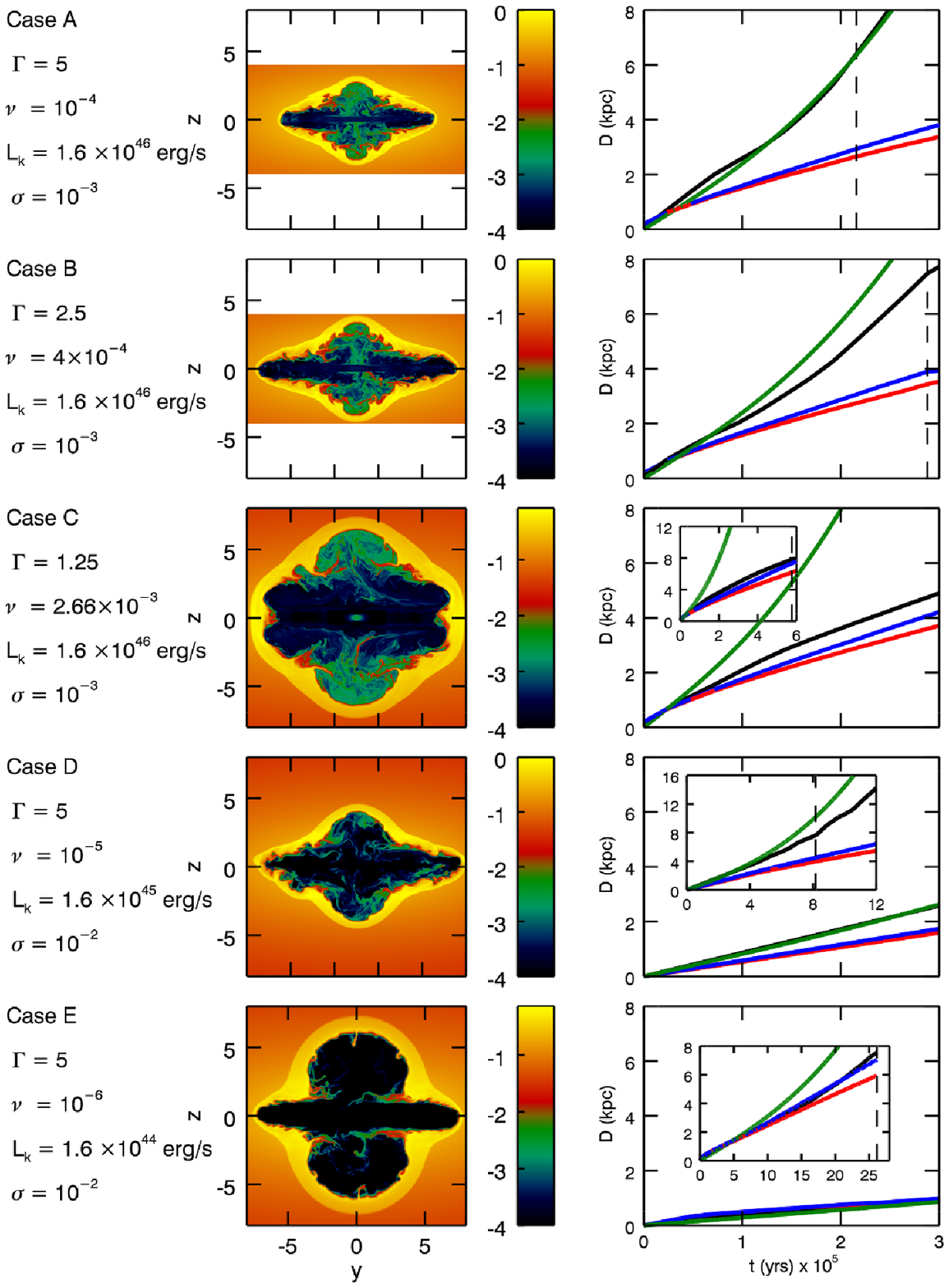} 
\caption{The  panels in the central column show 2D cuts of the logarithmic density distributions in the $yz$ plane for  cases A to E,  at the time when the cocoon has reached a size of approximately  $8$ kpc in the $y$ direction (the times of each snapshot is shown in the right panels by the vertical dashed lines).  The density is in units of $\rho_0$. The right panels show plots of the cocoon size along the three axes (black along $y$, blue along $z$ and red along $x$) as a function of time up to  $3 \times 10^5$ yrs, we recall that the $z$ direction corresponds to the minor axis, while the $x$ direction corresponds to the intermediate axis. The green lines show the theoretical estimate of the jet head position based on Eq. \ref{eq:vhead} Since the expansion of cases C, D and E is slower, the insets in the three bottom figures show the evolution for later times. The parameters for each case are given on the left.  
 }
   \label{fig:fig2}
\end{figure*}

\begin{figure*}[t!]
   \centering
   \includegraphics[width=15cm]{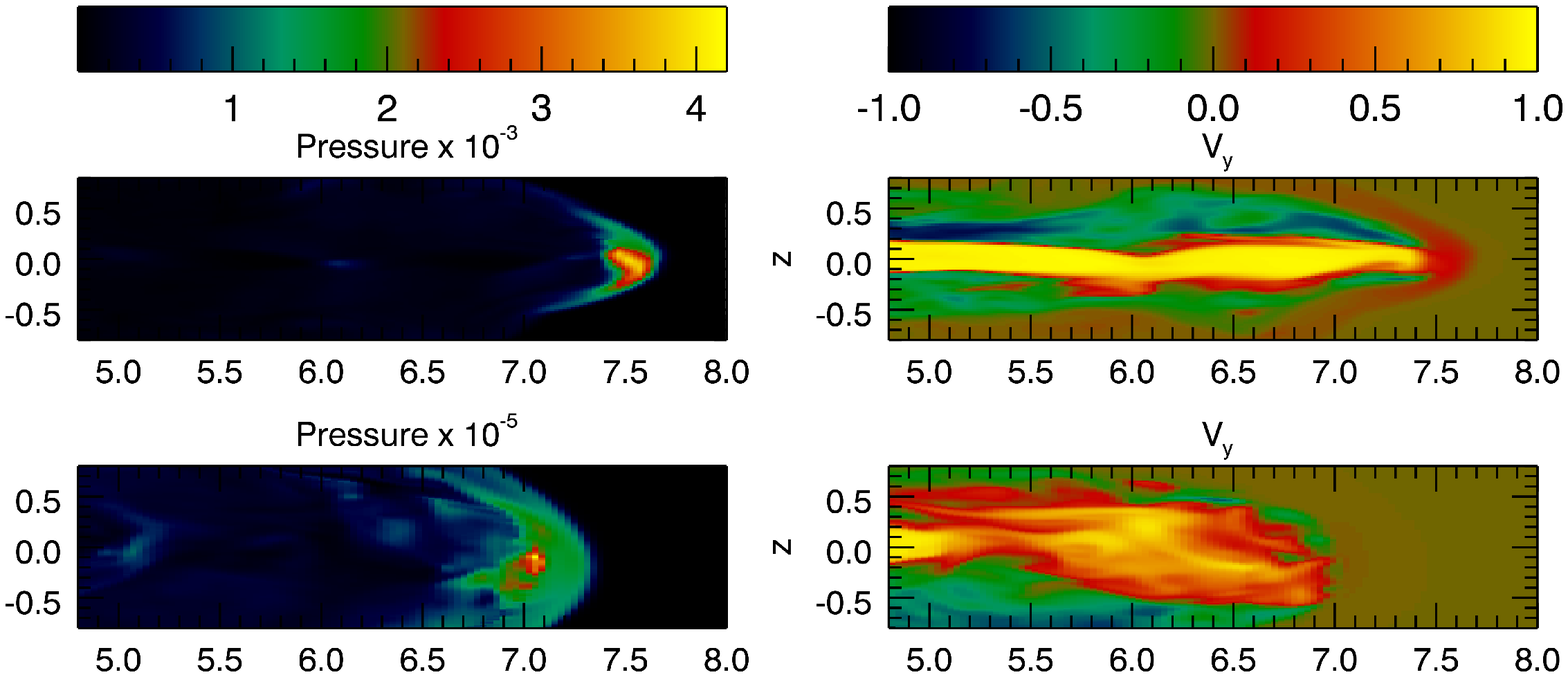} 
\caption{The left panels show the pressure distributions in the $yz$ plane, i.e. the plane defined by the major and minor axes, while the right panels show the velocity distributions in the same plane, respectively for cases A (top) and E (bottom). The pressure is in units of $\rho_0 c^2$, while the velocity is in units of $c$. We show only a part of the right jet in order to display the structure of the jet head. 
 }
   \label{fig:fig3}
\end{figure*}

In Fig. \ref{fig:fig2},  the panels in the central column  show, for cases A to E,   2D cuts in the $yz$ plane of the logarithmic density distribution at the time when the jets have reached a length of approximately  8 kpc.   We recall that the $z$ direction  corresponds to the minor axis of the triaxial distribution.  The figures give a first impression of the relative importance of lateral wings in 
the different cases. The formation of a winged structure depends on the lateral expansion velocity of the cocoon relative to the 
propagation velocity of the jet head.  A theoretical estimate of the latter can be derived by equating the momentum flux of the jet
and of the ambient medium in the frame of reference of the jet head. In the relativistic case one can derive 
the following expression \citep{Marti97}:
\begin{equation}
v_h = \frac{\gamma_j \sqrt{\eta}}{1+ \gamma_j \sqrt{\eta}} v_j \;,
\label{eq:vhead}
\end{equation}
where 
\begin{equation}
\eta = \frac{\rho_j h_j}{\rho_a h_a}
\end{equation}
and $\rho_j$, $\rho_a$, $h_j$, $h_a$ are respectively the jet and ambient densities and the jet and ambient specific enthalpies. The ambient density
decreases as the jet propagates outward according to Eq. (\ref{eq:density}) and, according to Eq. (\ref{eq:vhead}), the jet head should  accelerate. 
In the right column of Fig.  \ref{fig:fig2}, we can 
compare the actual jet head position as a function of time, represented by the black curves, with the theoretical estimates given by Eq. (\ref{eq:vhead}), represented by the green curves.  In the same figures we also show the size reached by the cocoon as a function of time, in the two transverse direction, the size along the $z$ direction is represented in blue, while the size along the $x$ direction is represented in red.  Considering cases A, B and C, the initial predicted jet head velocities are the same, since $\gamma_j^2 \rho_j$ is kept constant, and equal to about $2.1 \times 10^9$ cm s$^{-1}$.  The actual values are in very good agreement with this estimate. These three cases, according to Eq. (\ref{eq:vhead}) should behave in the same way, however, in reality, they show marked differences. In case A the jet head accelerates reaching a velocity of about $4.3 \times 10^9$ cm s$^{-1}$ when the jet extends up to 8 kpc. The agreement between the actual and estimated velocities is very good.
In case B we still observe acceleration and the velocity at 8 kpc is about $3.15\times 10^9$ cm s$^{-1}$, the comparison with the theoretical estimate shows an actual lower velocity, in fact 
the estimated velocity at 8 kpc should be as before $4.3 \times 10^9$ cm s$^{-1}$. This effect is much more pronounced in case C, where, instead of accelerating, the jet head decelerates, reaching a velocity of  about  $8.7 \times 10^8$ cm s$^{-1}$ at 8 kpc. According to Eq. (\ref{eq:vhead}), the jet head velocity, in our conditions, scales as $\sqrt{\nu}$ and therefore the initial values reduce to  $6.7 \times 10^8$ cm s$^{-1}$ in case D and to  $2.1 \times 10^8$ cm s$^{-1}$ in case E. In case D,  we still observe an acceleration of the jet head, with, however, a  velocity lower than expected (compare the black and green curves), with an actual value of about  $1.6\times 10^9$ cm s$^{-1}$. In case E  the deviation from the theoretical estimate is more evident and the  acceleration barely visible (compare the black and green curves), in fact the velocity at 8 kpc is only $3.13\times 10^8$ cm s$^{-1}$. Clearly  the jet reaches a  size of 8 kpc  at progressively later times in cases D and E.   The discrepancy between the actual and estimated velocity can be understood by considering that, in deriving Eq. (\ref{eq:vhead}), it is implicit the assumption that the two areas used for balancing the momentum fluxes are equal \citep{Lind89, Massaglia96}, in reality the area on which the jet deposits its momentum can increase leading  to a velocity lower than expected from Eq. (\ref{eq:vhead}).   The increase in the area can occur either because of an expansion of the working surface, particularly evident in case C, or because of instabilities that can fragment or induce wiggling of the jet head. In Figure \ref{fig:fig3} we compare the structures of the jet head for case A, in which the actual head velocity perfectly agrees with the theoretical estimate, and for case E, in which the actual head velocity is well below the theoretical estimate. The top panels refer to case A, while the bottom panels refer to case E, left panels show a cut of the pressure distributions in the plane $yz$,  i.e. the plane defined by the major and minor axes, while right panels show a cut of the longitudinal velocity distributions in the same plane. We can see that while in case A all the jet thrust is concentrated on a section which is comparable to the jet width, in case E the jet close to the head appears to have widened and the thrust is distributed on a larger area. This is consistent with the discussion made above  and the effect appear to become more pronounced  as we decrease the density ratio.

\begin{figure}[h!]
   \centering
   \includegraphics[width=8cm]{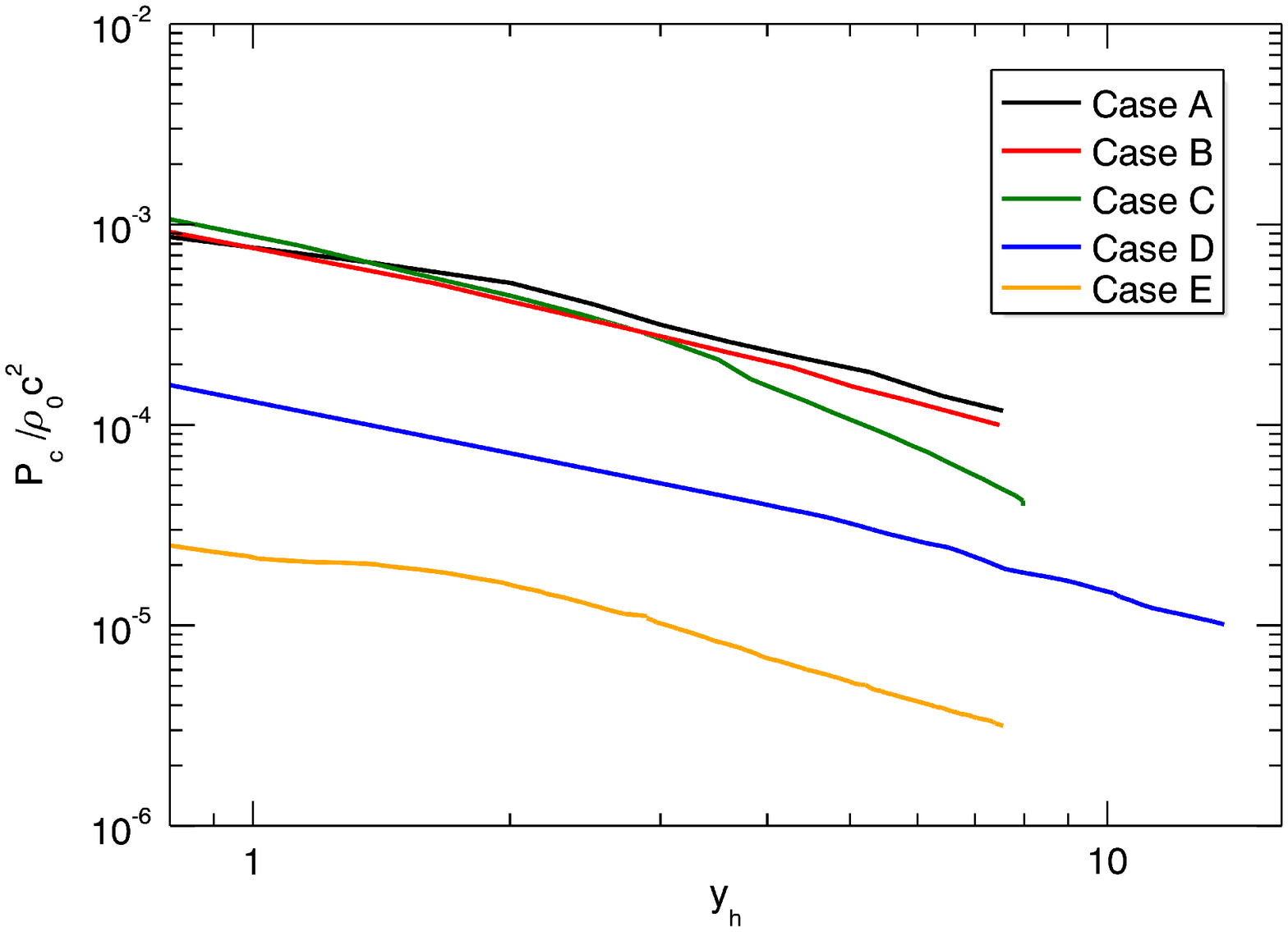} 
\caption{Plot of the average pressure (averaged over the cocoon volume) as a function of jet length for all cases. 
 }
   \label{fig:cocoon_pressure}
\end{figure}

Concerning the wing expansion, their velocity can be estimated as \citep{Begelman89}:
\begin{equation}
v_w = \sqrt{\frac{p_c}{\rho_a}}
\end{equation}
where $p_c$ is the average cocoon pressure, whose evolution as a function  of the jet length is represented in Fig.  \ref{fig:cocoon_pressure}, for all cases. We see that the cocoon pressure decreases as the system evolves and its size increases. The wings expansion velocity which depends on the ratio between $p_c$ and $\rho_a$, will increase or decrease depending on which between pressure and density decreases faster. Our results displayed  in the right column of Fig. \ref{fig:fig2} show that the lateral expansion velocities for all cases remain approximately constant in time (blue and red curves), the decrease in the cocoon pressure is then almost balanced by the decrease in the external density. 

In cases A, B and C, the cocoon pressure has approximately the same value and in fact the lateral expansion velocities are almost the same for all these cases,  approximately $10^9$cm s$^{-1}$, with the velocity along the $x$ direction about $10\%$ lower than the velocity along the $z$ direction. Notice then that the higher wing prominence in case C is due to a lower jet head velocity and not to a higher lateral velocity. The results for cases A, B and C show that for getting a winged source at high power the jets have to be subrelativistic. For relativistic jets we can get X-shape morphology only decreasing their power. 

\begin{figure}[h!]
   \centering
   \includegraphics[width=8cm]{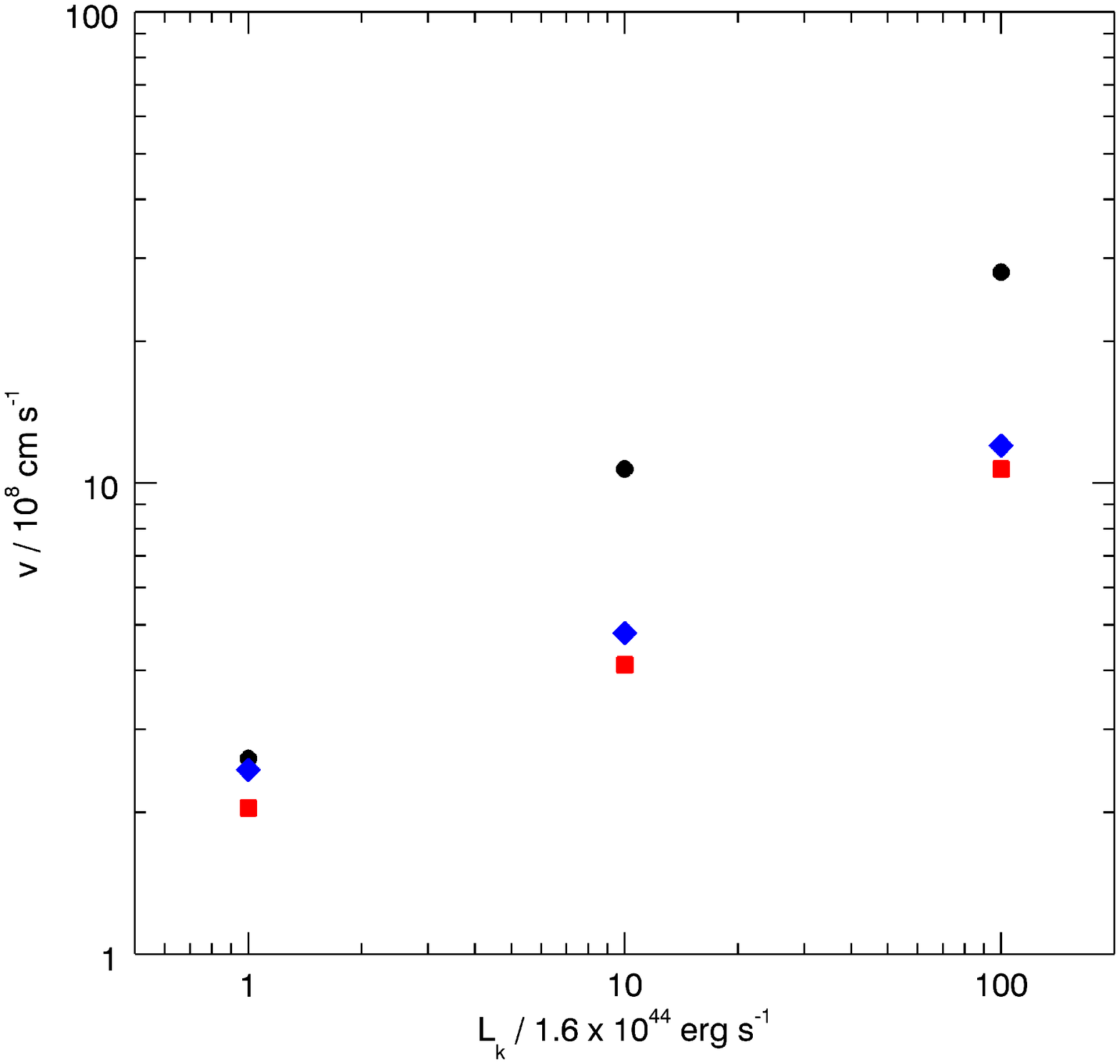} 
\caption{Expansion velocities along the three axes as a function of the jet power. Circles are for the $y$ direction, diamonds for the $z$ direction and squares for the $x$ direction.
 }
   \label{fig:fig5}
\end{figure}

The decrease of the jet power leads to a decrease of the cocoon pressure, as
shown in Fig. \ref{fig:cocoon_pressure}, and therefore to a slowing down of
the wing expansion. The lateral velocities reduce to about $5 \times 10^8$ cm
s$^{-1}$ in case D and to about $2.5 \times 10^8$ cm s$^{-1}$ in case E.
However, as seen above, also the jet head velocity reduces with the jet power
and the wing prominence depends on which reduces faster. In
Fig. \ref{fig:fig5} we compare, for cases A, D and E, the average expansion
velocities along the three axes as a function of jet power. Circles represent
the jet head velocity, while diamonds and squares represent the lateral
expansion velocities respectively along the $z$ and $x$ directions. The figure
shows that the jet head velocity decreases with jet power faster than the
lateral velocities and moreover the expansion velocity along the minor axis
$z$ decreases slower than the corresponding velocity along the intermediate $x$
axis.  Therefore, in conclusion, only low power relativistic jets can give
origin to X-shaped morphologies, while high power jets have to be
subrelativistic, as shown by cases A, B and C discussed above. Based on the
comparison between the velocities along the $z$ and $x$ axes, shown in Fig
\ref{fig:fig5}, we expect that, increasing the ellipticity, the decrease of
the lateral expansion velocity along the minor axis will be lower and the
critical power, under which a winged morphology can be obtained, will depend
on the ellipticity of the gas distribution, with a higher critical power for
higher ellipticity.
    
\subsection{Case F: The effects of misalignment between jet and major axis}

\begin{figure*}[t!]
   \centering
   \includegraphics[width=17cm]{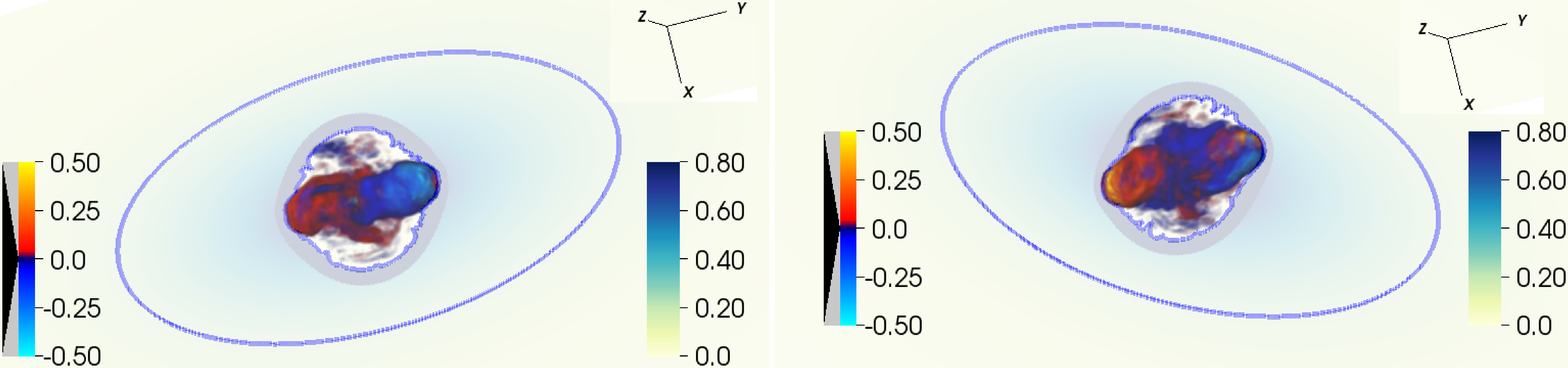} 
   
   \vspace{0.3cm}
   
     \includegraphics[width=17cm]{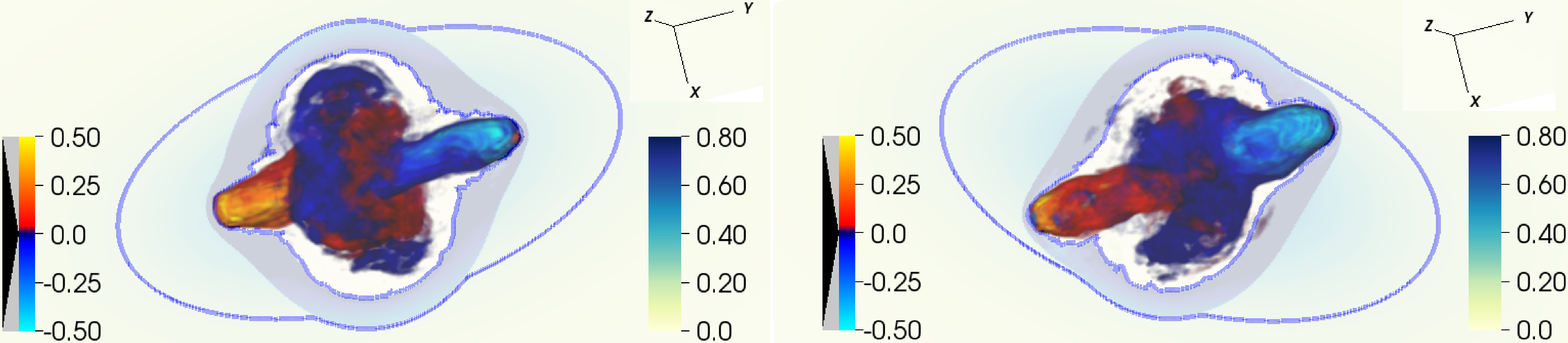} 
     
   \vspace{0.3cm}

     \includegraphics[width=17cm]{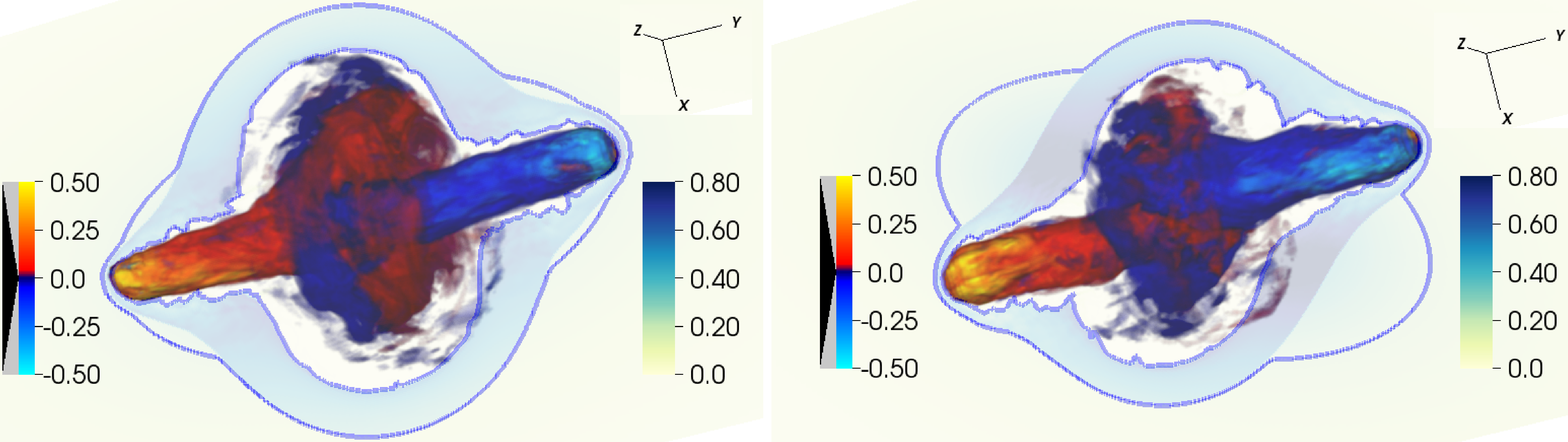} 

\caption{Evolution of  cases E and F. The three panels on the left are for case E at times respectively $9.6 \times 10^5$ yrs, $1.8  \times 10^6$ yrs,  and $2.6 \times 10^6$ yrs,, from top to bottom, while the three panels on the right are for case F at times respectively $9.6 \times 10^5$ yrs, $1.6  \times 10^6$ yrs, , $2.3  \times 10^6$ yrs, from top to bottom. 
The distribution of $y$ velocity component   is represented by a volume rendering, with blue colors representing negative values and red-yellow colors representing positive values.  Superimposed, we also show a cut of the density distribution in the $x-y$ plane, while the blue lines are density iso-contours at $\rho = 0.2$cm$^{-3}$ on the same plane.
}
   \label{fig:fig3d}
\end{figure*}

\begin{figure*}[t!]
   \centering
   \includegraphics[width=12cm]{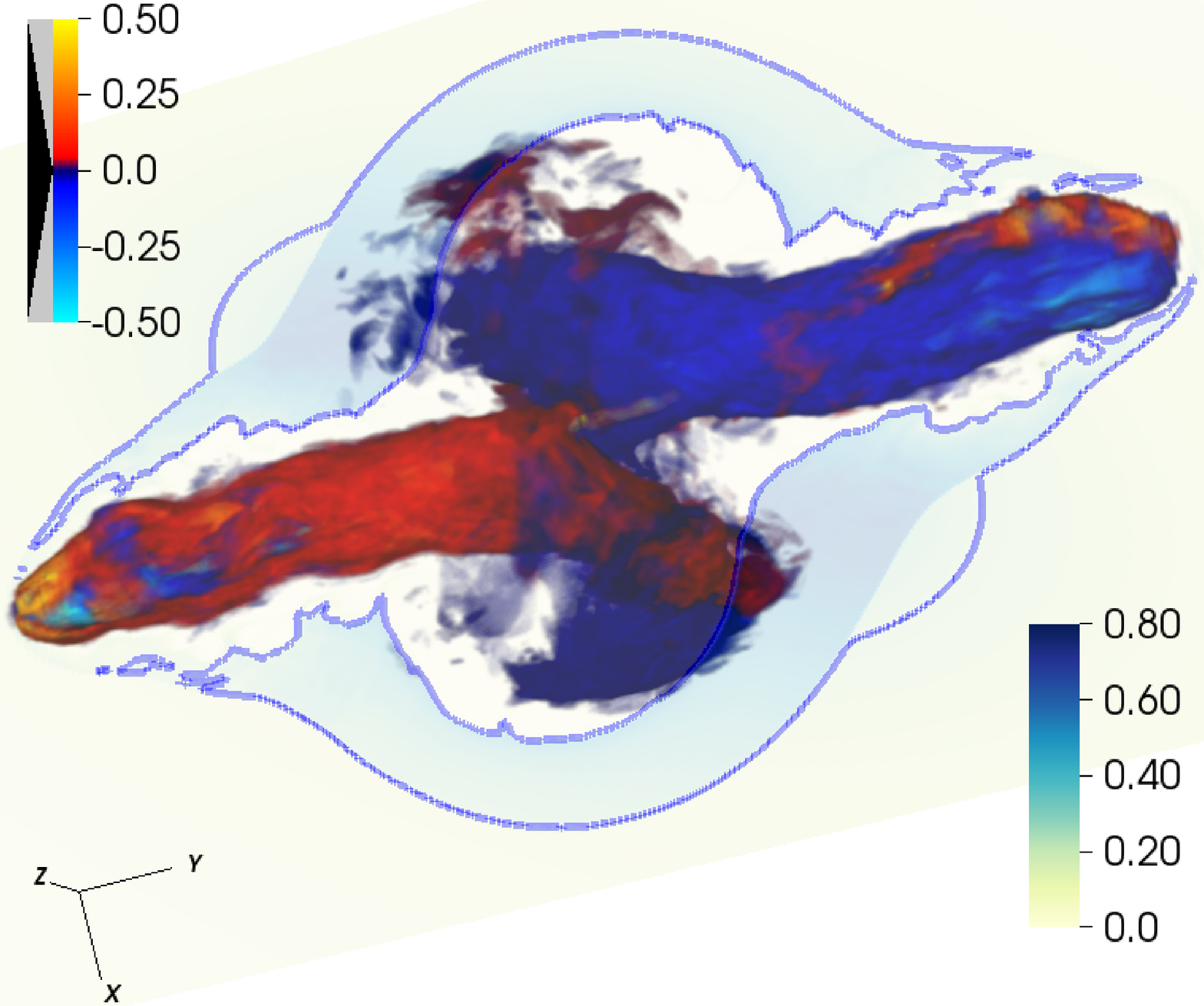} 
\caption{Backflow structure for case F at $t = 3.2 \times 10^6$ yrs. The representation is the same as in Fig. \ref{fig:fig3d}
 }
   \label{fig:fig3d2}
\end{figure*}

In this subsection we will describe the evolution of Case F, which has the
same parameters as case E, but has a misalignment of $30^o$ between the jet
axis and the major axis of the density distribution and which we followed for
a somewhat longer time, until it reached a jet length of about 12 kpc at $t
\sim 3.6\times10^6$yrs. The most favourable condition for the formation of the
wings is when the jet is aligned with the major axis of the galaxy, with case
F we want then explore whether the X-shape morphology can develop also in a
less favourable condition. Fig. \ref{fig:fig3d} compares the evolution of case
E and case F by showing three-dimensional volume renderings of the
distribution of the $y$ component of the velocity, with blue colors
representing negative values and red-yellow colors representing positive
values. The renderings show in particular the structure of the backflows,
i.e. of the jet material that after crossing the jet terminal shock flows back
towards the jet origin. Superimposed, we also show a cut of the density
distribution in the $x-y$ plane, while the blue lines are density iso-contours
at $\rho = 0.2$cm$^{-3}$ on the same plane. The density distributions and the
iso-contour lines allow to follow how the initial ellipsoidal density structure
is progressively modified by the inflating cocoon.  The panels on the left are
for case E, while the panels on the right are for case F. The three panels on
each column are for three different times showing the cocoon
evolution. Finally in Fig. \ref{fig:fig3d2} we show only the structure of case
F at a later time, for which we don't have the corresponding image for case E,
since, for this case we stopped the simulation at an earlier time.  The figure
shows how the cocoon progressively inflates expanding in the direction of the
minor axis of the density distribution, which in case F is not perpendicular
to the jet axis.  As shown in the figure, the times at which case F reaches
the same length as case E are somewhat earlier, indicating a slightly larger
jet head velocity. In fact the average velocity for case F is $3.26 \times
10^8$ cm s$^{-1}$ to be compared to $2.8 \times 10^8$ cm s$^{-1}$. The lateral
expansion velocity is, on the contrary, slightly lower in case F than in case
E. The initial density distribution, that at the beginning strongly influences
the cocoon evolution, is completely modified by the jet propagation, which
inflates a cavity and forms a compressed shell at the border of the cavity.
In addition we can follow the interaction of the two backflows, with one of
them always prevailing on the other and breaking, in case E, the initial
symmetry. In case F the jets form non symmetric backflows that bend towards
the minor axis of the initial density distribution flowing up in the wings, as
it can be seen in Fig. \ref{fig:fig3d2}.  In case F the wing axis form an
angle of about $70^\circ$ with the initial jet axis, almost constant with
time, and slightly larger than the angle of $60^\circ$ formed by the jet axis
with the minor axis of the density distribution.

\begin{figure*}[ht!]
   \centering
   \includegraphics[width=15cm]{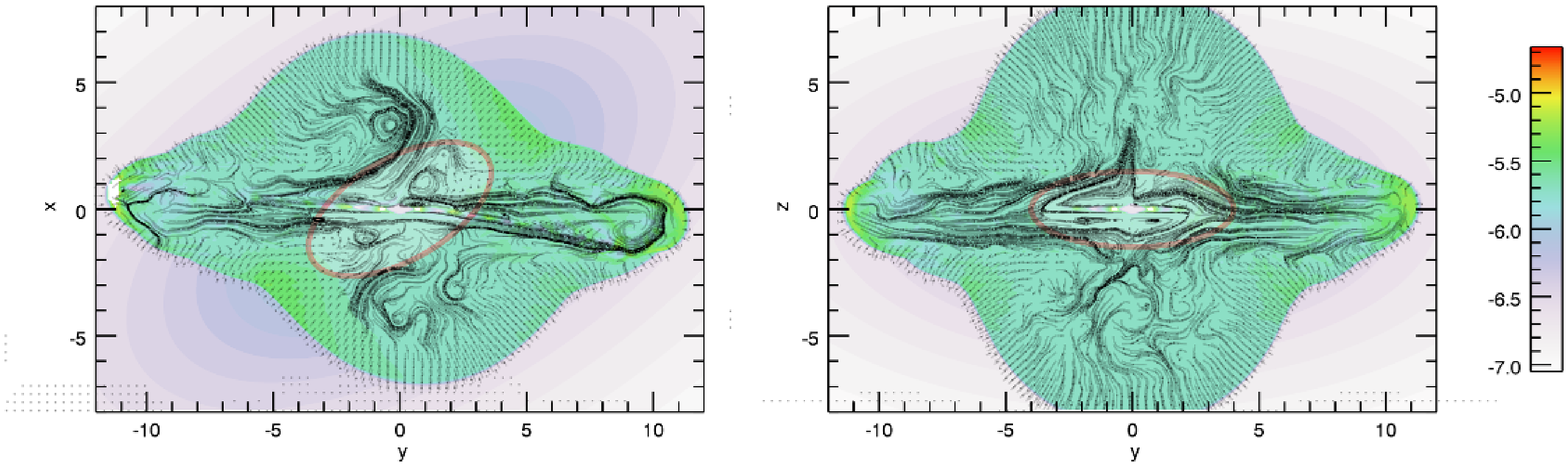} 
\caption{2D cuts of the pressure distribution in the $yx$ (on the left) and $yz$ (on the right) planes for case F, at the time when the cocoon has reached a size of approximately $24$ kpc in the $y$ direction. Superimposed we show also the velocity field. The red ellipses represents equicontours of the initial density distribution and indicate the relative orientation of the jet with respect to the density distribution.
 }
   \label{fig:fig8}
\end{figure*}

Fig. \ref{fig:fig8} shows the distribution of pressure in the two planes $(xy)$ and $(yz)$, when the radio source has reached the maximal size at $t \sim 3.2 \times 10^6$ yrs, with superimposed the velocity field. The red ellipses represent an equicontour of the initial density distribution.  In the two  panels we show respectively cuts in the $x-y$ plane (left panel) and in the $y-z$ plane (right panel).  The two panels allow to observe the effect of the misalignment on the final geometrical structure, already evident in Fig. \ref{fig:fig3d}.    In addition we can see that the cocoon is still strongly overpressured with respect to the ambient.  Looking at the velocity field, whose streamlines are visible in the same figure, we can observe the presence of flows rising in the wings. 

The conclusion of the analysis of case F is that the prominence of wings is lower than in case E, as could be expected, since the jet direction has been moved away from the most favourable condition of perfect alignment with the major axis of the initial density distribution. This inclination of $30^o$, however, is still  favourable for the formation of a winged source.

Summarizing, we showed that jets with a power larger than $10^{45}$ erg
s$^{-1}$ have to be non relativistic in order to form a winged structure,
while relativistic jets can form a winged structure only if they have $L_j
\lesssim 10^{44}$ erg s$^{-1}$. \cite{birzan04} estimated the jet power
 from the observations of the X-ray cavities inflated by radio
AGN. The ratio between $L_j$ and the radio power at 1.4 GHz ($P_{1.4}$)
ranges from $\sim 3$ to $\sim 300$ and this leads to an estimate of $P_{1.4}
\lesssim 3 \times 10^{43}$ erg s$^{-1}$ for the maximum radio luminosity of
winged sources; this value compares favourably with the power distribution of
X-shaped radio sources, that reaches $P_{1.4} \sim 2 \times 10^{43}$ erg
s$^{-1}$ \citep{Gillone16}.

\subsection{Magnetic field structure}

\begin{figure*}[t!]
   \centering
   \includegraphics[width=6cm]{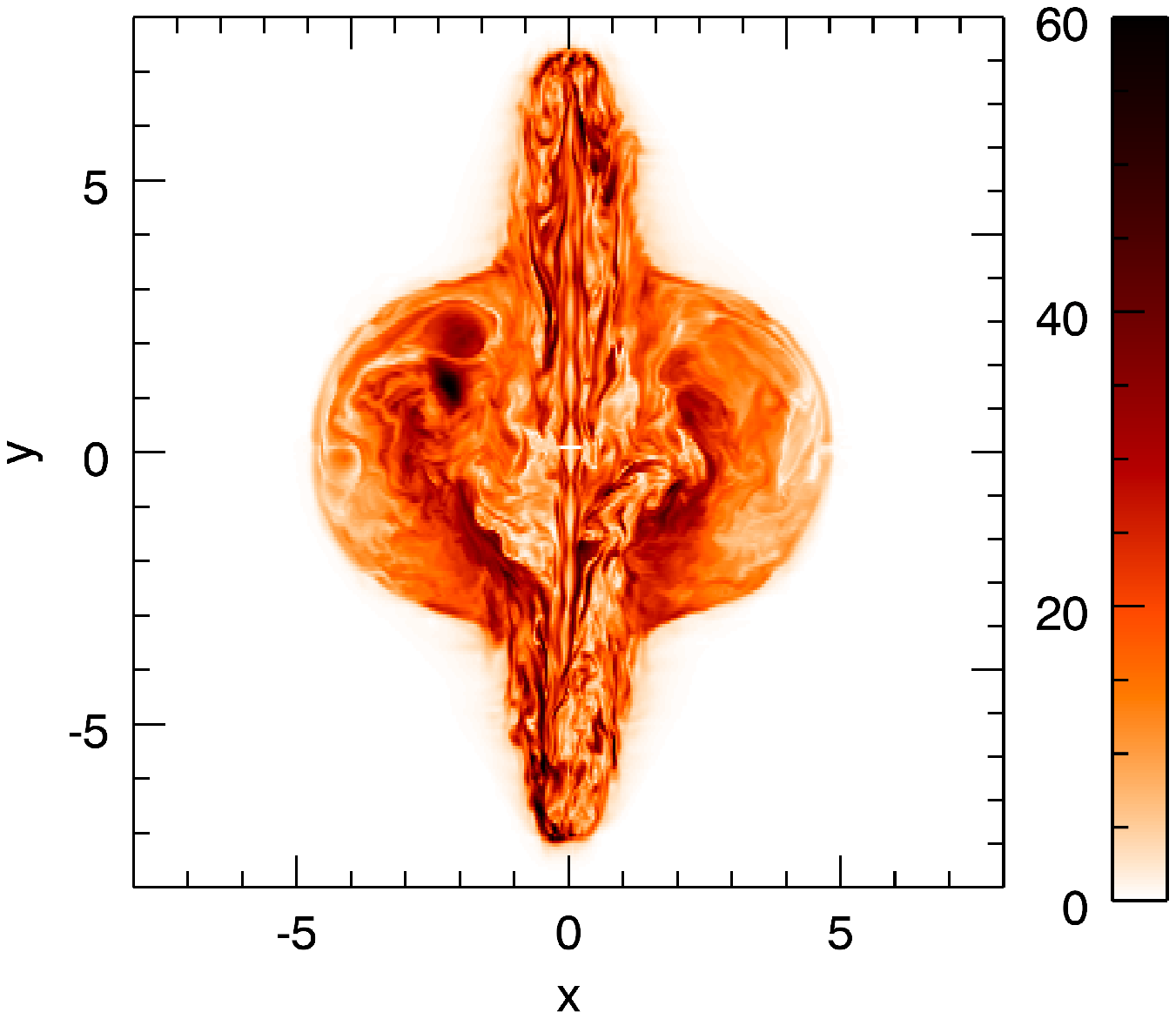} 
      \includegraphics[width=6cm]{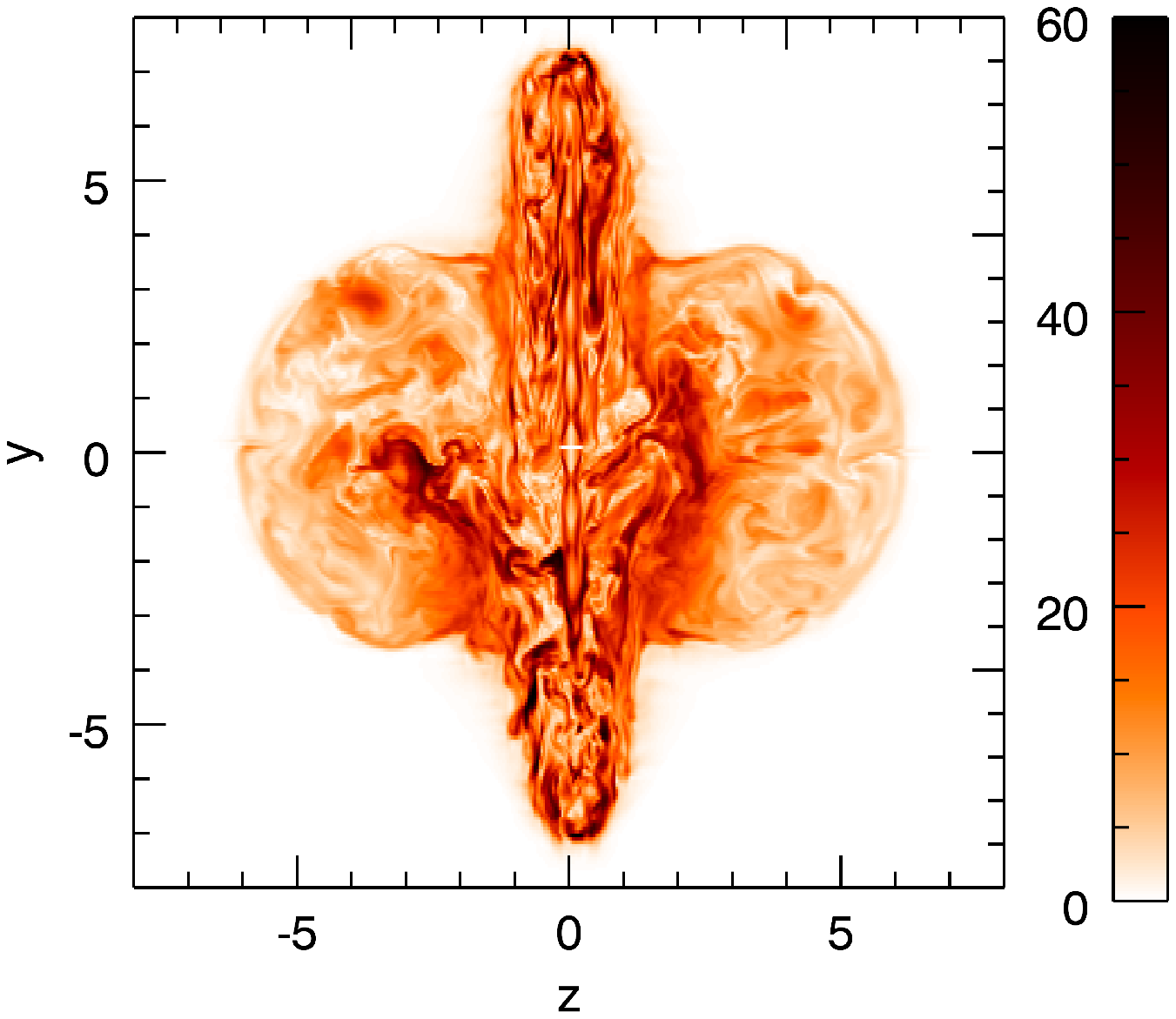} 
\caption{Two-dimensional cuts of the distribution of the  magnetic field strength. The left panel is in the $xy$ plane, the right panel is in the $yz$ plane. The field strength is expressed in microGauss, assuming for the basic physical units $r_c$ and $\rho_0$ their reference values. 
 }
   \label{fig:fig9}
\end{figure*}

\begin{figure}[h!]
   \centering
   \includegraphics[width=8cm]{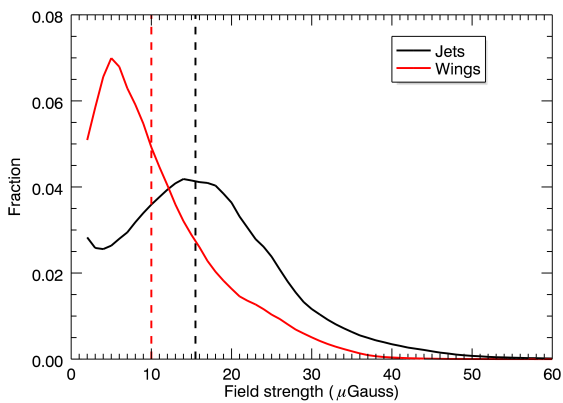} 
\caption{Distribution functions of the magnetic field strength in the jet (black curve) and in the wings (red curve). The dashed lines indicate the medians of the distributions.
 }
   \label{fig:fig10}
\end{figure}

\begin{figure}[htbp]
   \centering
   \includegraphics[width=8cm]{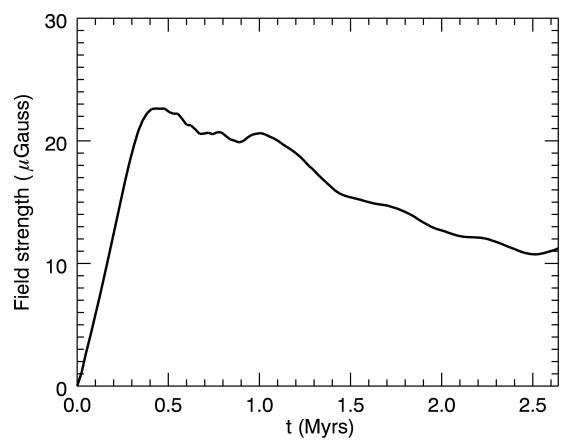} 
\caption{Plot of the average magnetic field as a function of time for case E.
 }
   \label{fig:fig11}
\end{figure}

An important question to be analysed is whether the flows rising in the wings
are able to carry the magnetic field necessary for sustaining the synchrotron
radiation. In order to answer this question, we first examine in
Fig. \ref{fig:fig9} the distribution of the magnetic field strength, for case
E at time $t = 2.6 \times 10^6$ yrs, in the $xy$ and $yz$ planes,
corresponding, respectively, to the planes of the major and minor axes ($xy$,
left panel) and of the major and intermediate axes ($yz$, right panel). The
field strength is expressed in microgauss and is computed by assuming for the
basic physical units $r_c$ and $\rho_0$ their reference values, recalling that
the field strength scales as $\rho_0^{1/2}$. From the figure we see that the
field strength in the jets reaches maximum values between $30 \mu$G and $60
\mu$G, similar values can be also found in the wings in the $xy$ plane and in
the lower part of the wings in the $yz$ plane, while the upper part of the
wings in the $yz$ plane shows lower values.  One question that immediately
arises is whether the magnetic field becomes dynamically important, we have
then examined the distribution of the local values of the plasma $\beta$
representing the ratio between thermal and magnetic pressure and we have seen
that there are only a few points where it reaches values of the order of
unity, while for the most part it has values much larger than unity and
therefore it is generally dynamically unimportant. A more quantitative
comparison between the magnetic field strength in the jets and in the wings is
provided by Fig. \ref{fig:fig10} where we plot the distribution functions of
the magnetic field strength in the jet and in the wings. The jet is defined as
a cylinder of radius $0.3 r_c$ with its axis aligned with the $y$ axis, while
the wings are defined as the region external to this cylinder.  From the
figure we see that the maximum of the distribution is about $15 \mu$G in the
jets and about $5 \mu$G in the wings, the medians of the distributions are
however much closer, since they are again $15 \mu$G in the jets, but about $10
\mu$G in the wings. This means that half of the volume in the jets has field
strength larger than $15 \mu$G, while in the wings, in order to get half of
the volume, we have to lower the threshold to $10 \mu$G. Regions with field
strength larger than $30 \mu$G have very small filling factors, of about $7
\%$ in the jets and $1.5 \%$ in the wings.  The distribution of magnetic field
described above represents a snapshot of the evolution at $t = 2.6 \times
10^6$ yrs, at the end of the simulations, if we look at the complete temporal
evolution, we can see that the spatial distribution and the distribution
function of the field strength do not change much in time, however the average
field strength decreases with time as represented in Fig. \ref{fig:fig11}. The
value of the field depends of course on the value of $\sigma$, representing
the ratio of the Poynting flux to the matter energy flux, in this case $\sigma
= 10^{-2}$.  Note that the values of the resulting field amplitude are quite consistent with observations.   Increasing $\sigma$ we expect that the magnetic field scale as
$\sqrt{\sigma}$ until it reaches values for which it becomes dynamically important.  The regime in which the magnetic field becomes dynamically important is not examined in this paper, however it could be of interest for further investigations.

\subsection{Synthetic radiomaps}

\begin{figure*}[htbp]
   \centering
\includegraphics[width=8cm]{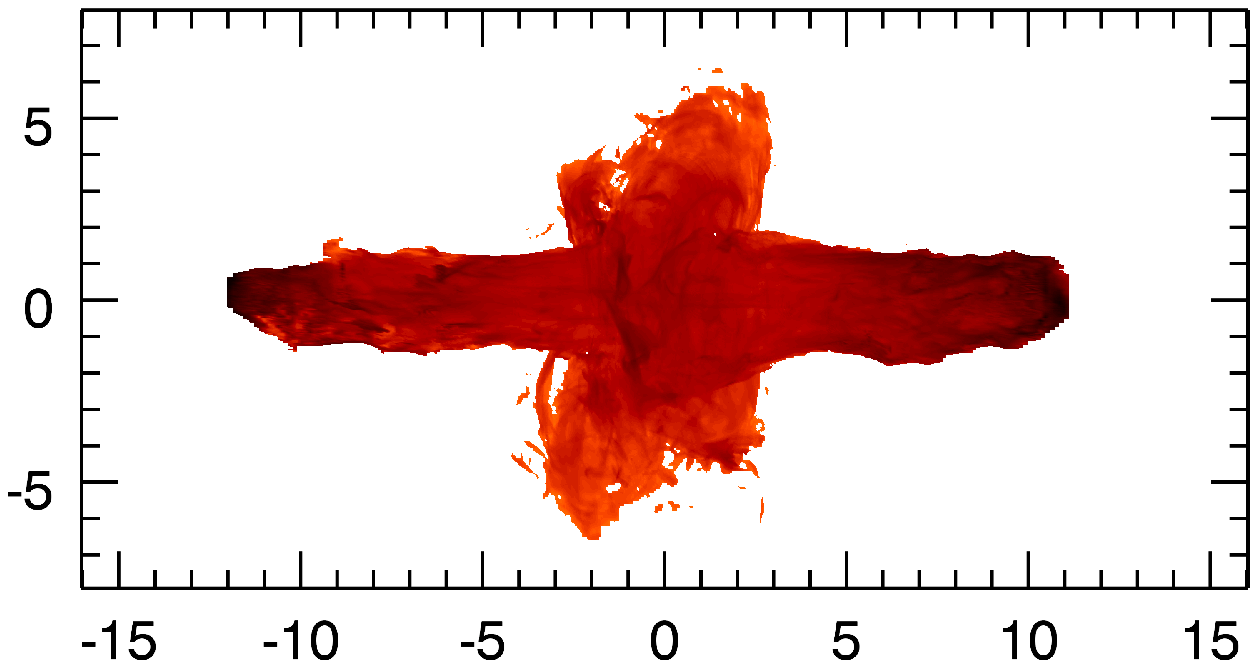} 
\includegraphics[width=8cm]{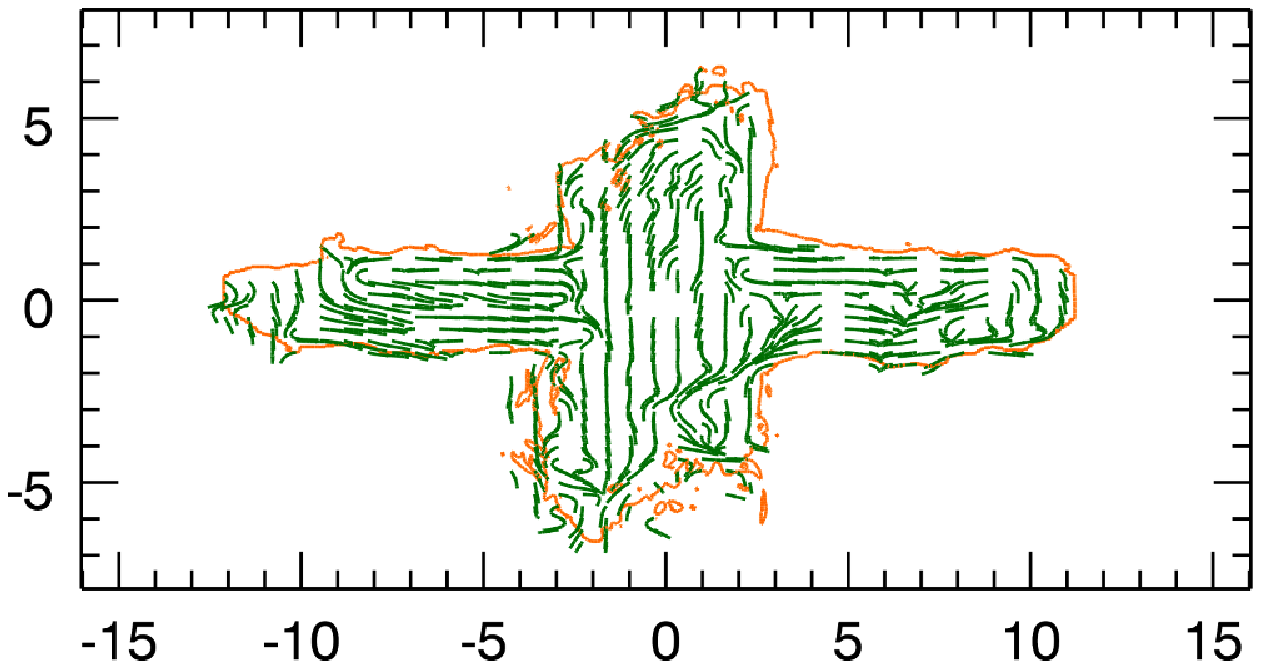} 
 \includegraphics[width=8cm]{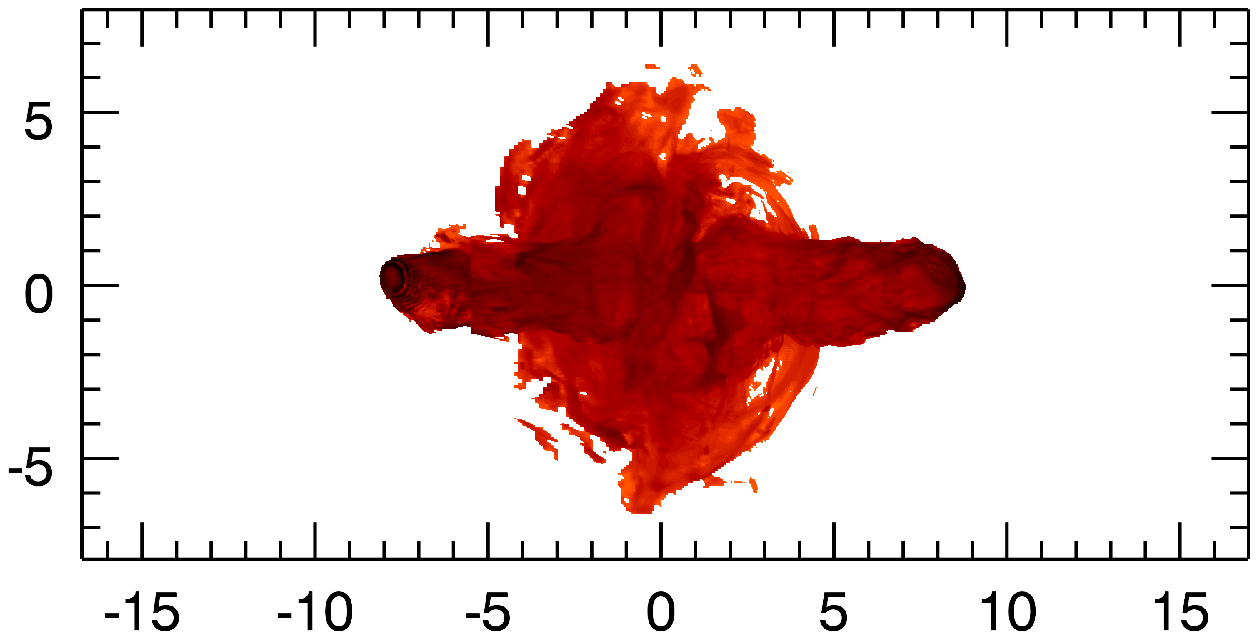} 
  \includegraphics[width=8cm]{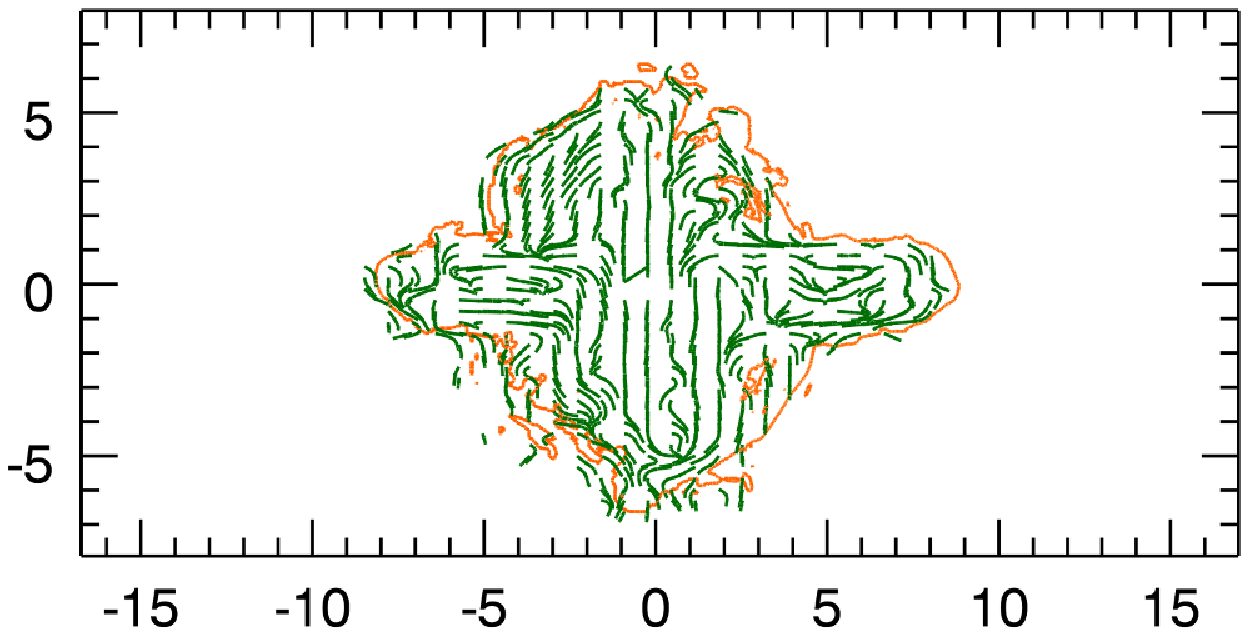} 
 \includegraphics[width=8cm]{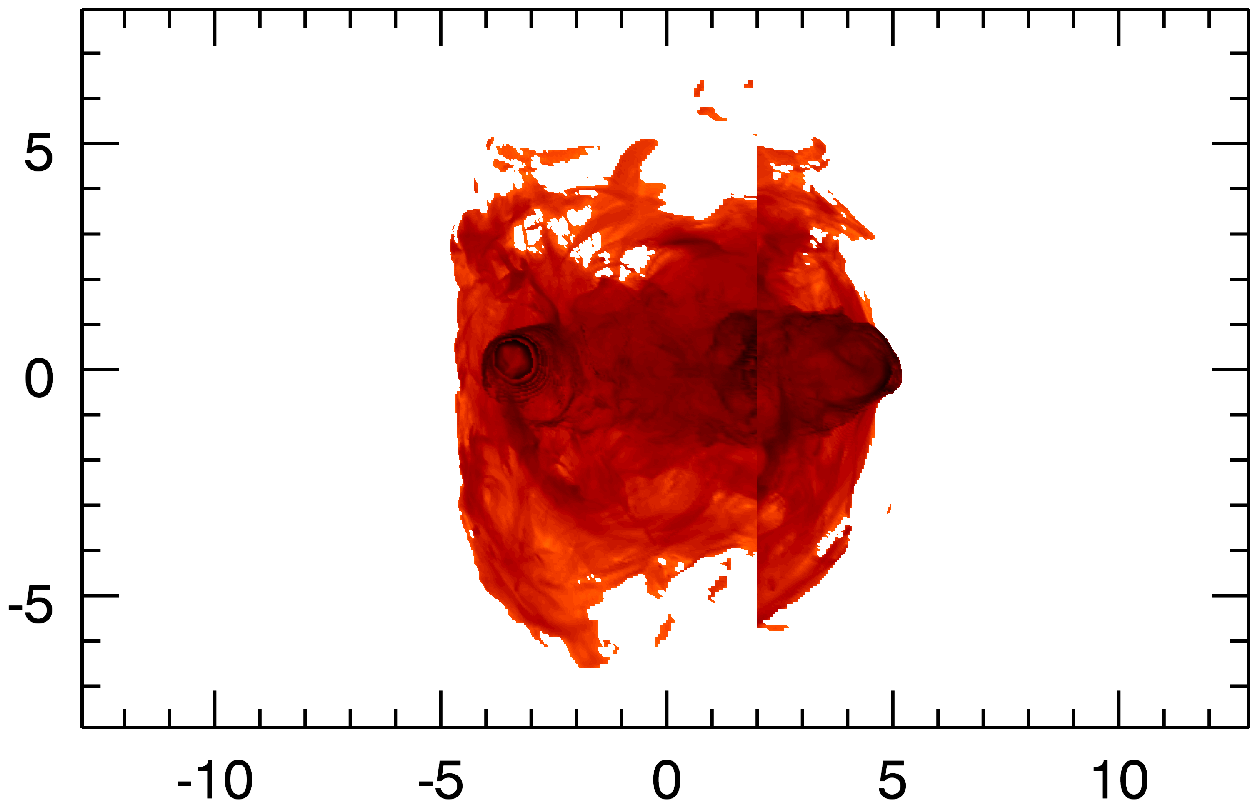} 
  \includegraphics[width=8cm]{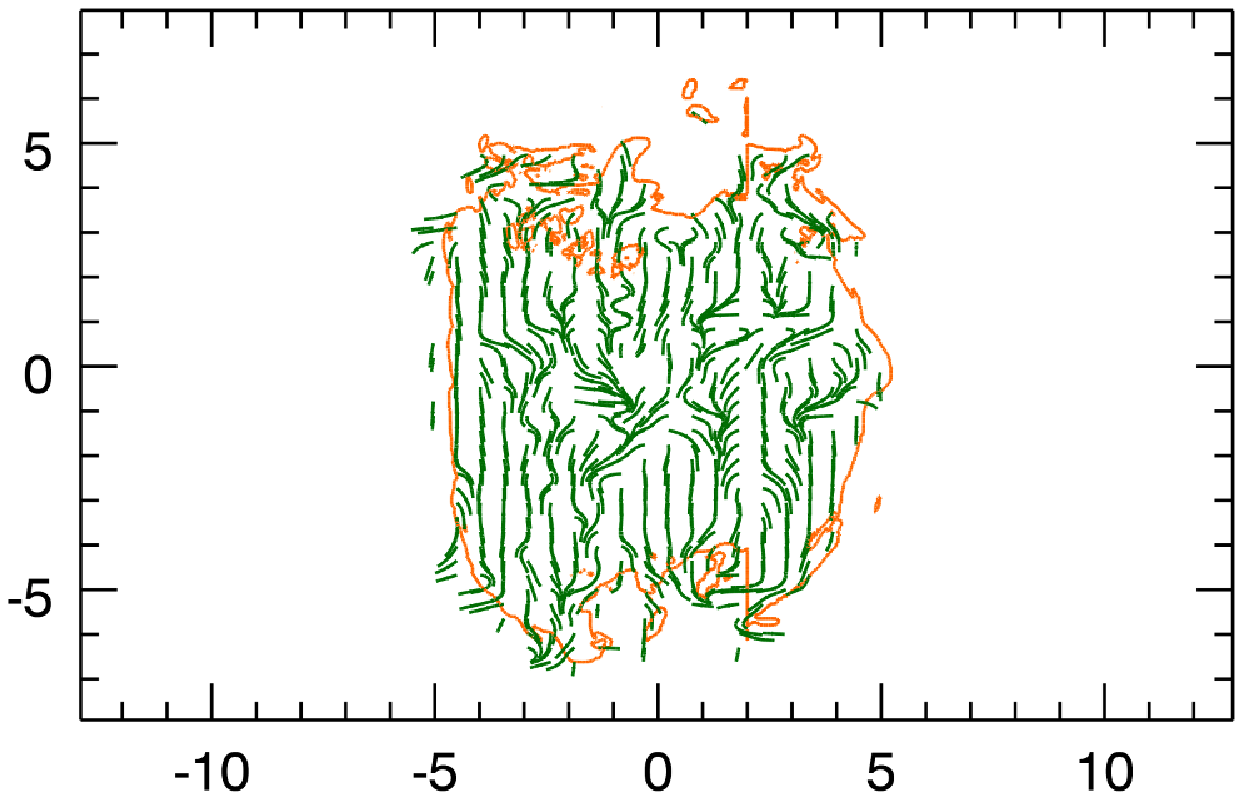} 

\caption{Synthetic radio surface brightness and polarization maps for different line of sights, for case F. The panels on 
the left show the brightness maps, with the line of sight in the $xy$ plane, while those on the right show the 
polarization maps (the magnetic field direction is plotted).  For the top panels the line of sight makes an angle of  
$90^\circ$  with the jet direction, in the middle the angle is $45^\circ$ and at the bottom  the angle is $20^\circ$. The 
brightness maps have a logarithmic scaling and cover a range of four order of magnitude.
 }
   \label{fig:fig12}
\end{figure*}

In Fig. \ref{fig:fig12} we display synthetic surface brightness and
polarization maps for different viewing angles. The calculations of the
surface brightness and of the polarization are described in the appendix \ref{ap:maps}. In
the left panels we show the brightness maps, while in the right panels we show
the polarization maps, plotting the magnetic field direction. The line of
sight is in the $xy$ plane and from top to bottom makes with the jet direction
angles respectively of $90^\circ, 45^\circ$ and $20^\circ$.  Decreasing the
viewing angle, the size of the wings becomes more prominent as the projected
jet length decreases, until, at the smallest viewing angles, the main lobes
are projected onto the wings.

\begin{figure*}[htbp]
   \centering
   \includegraphics[width=4cm]{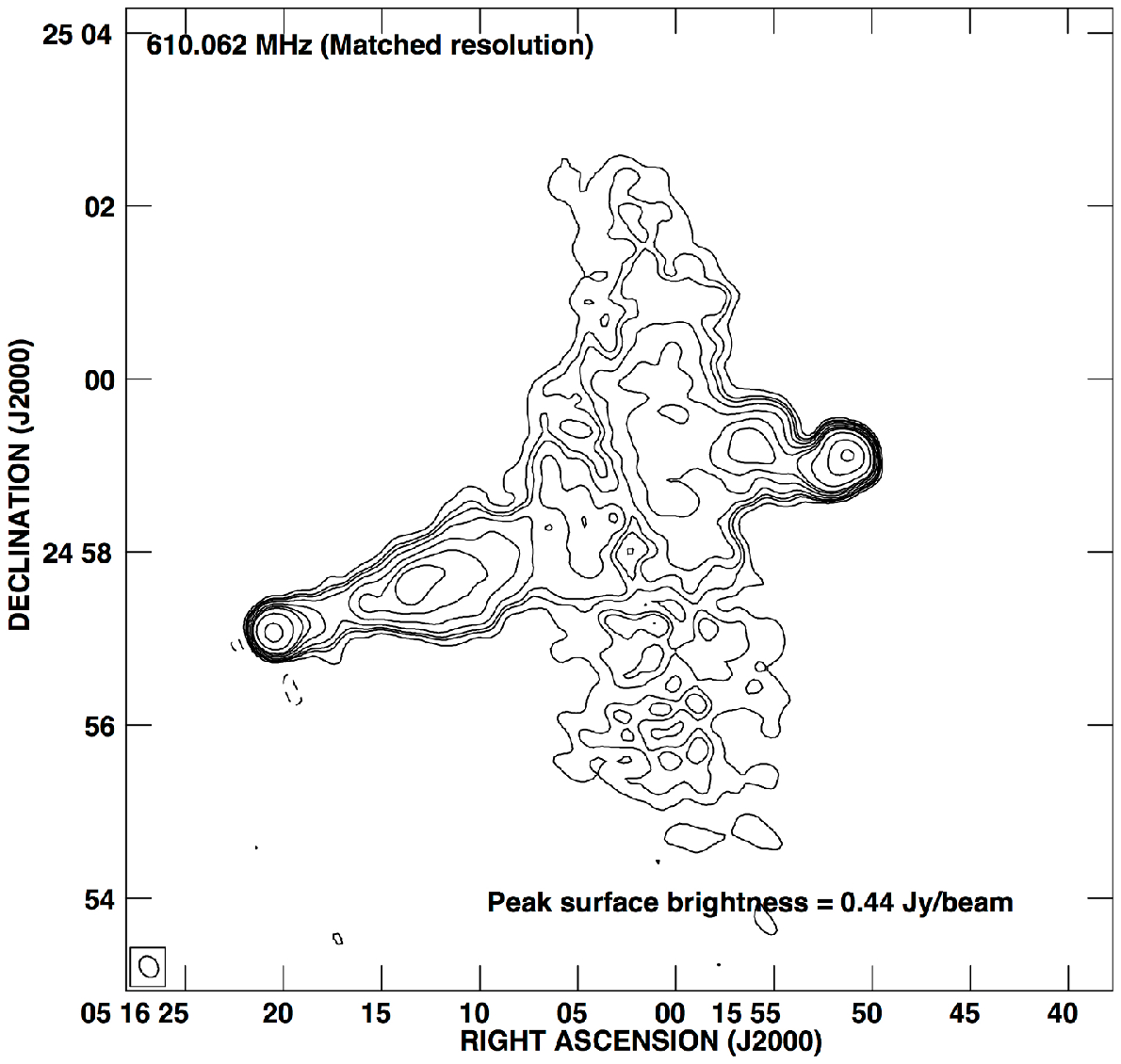} 
   \includegraphics[width=4cm]{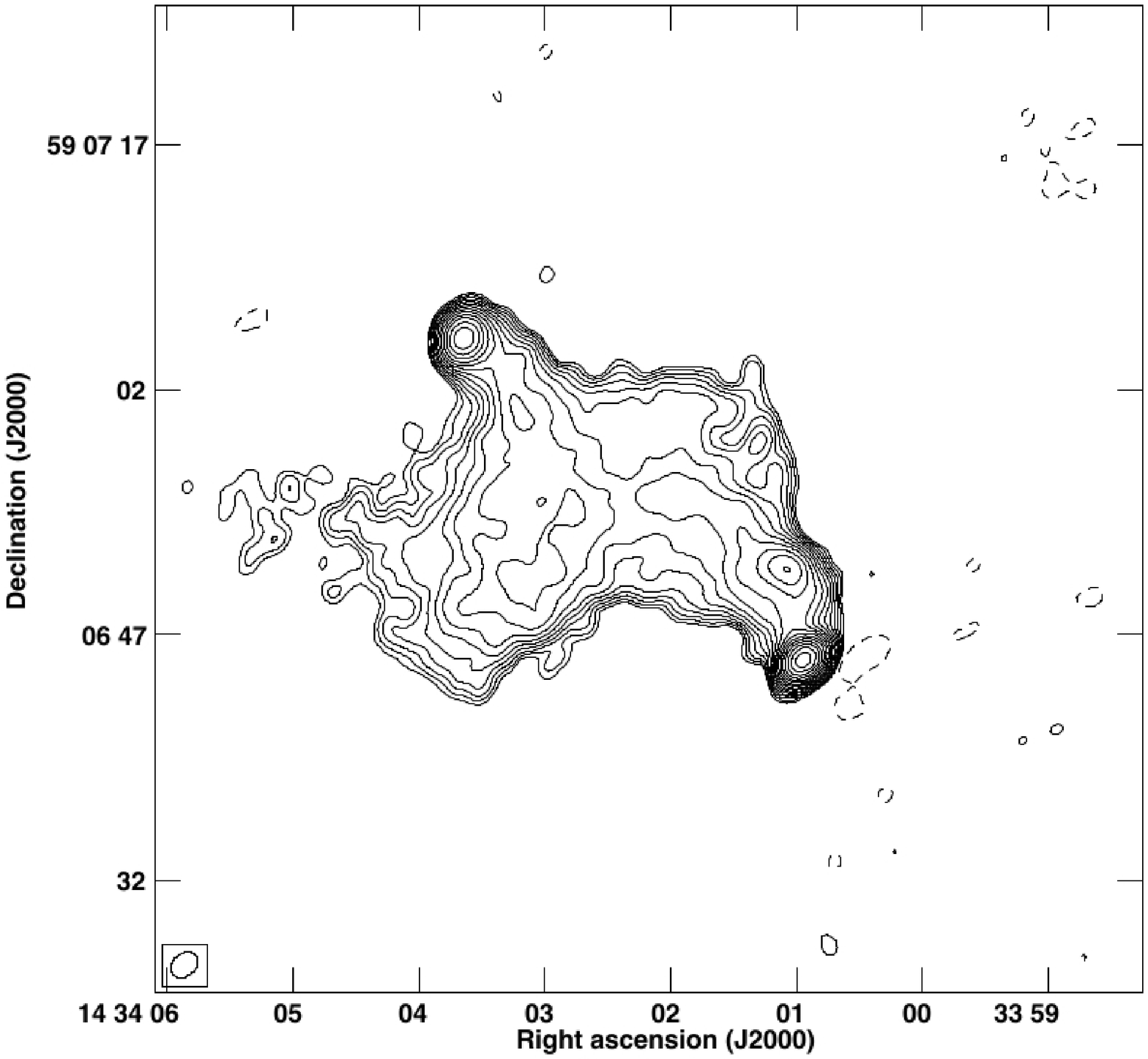} 
   \includegraphics[width=4cm]{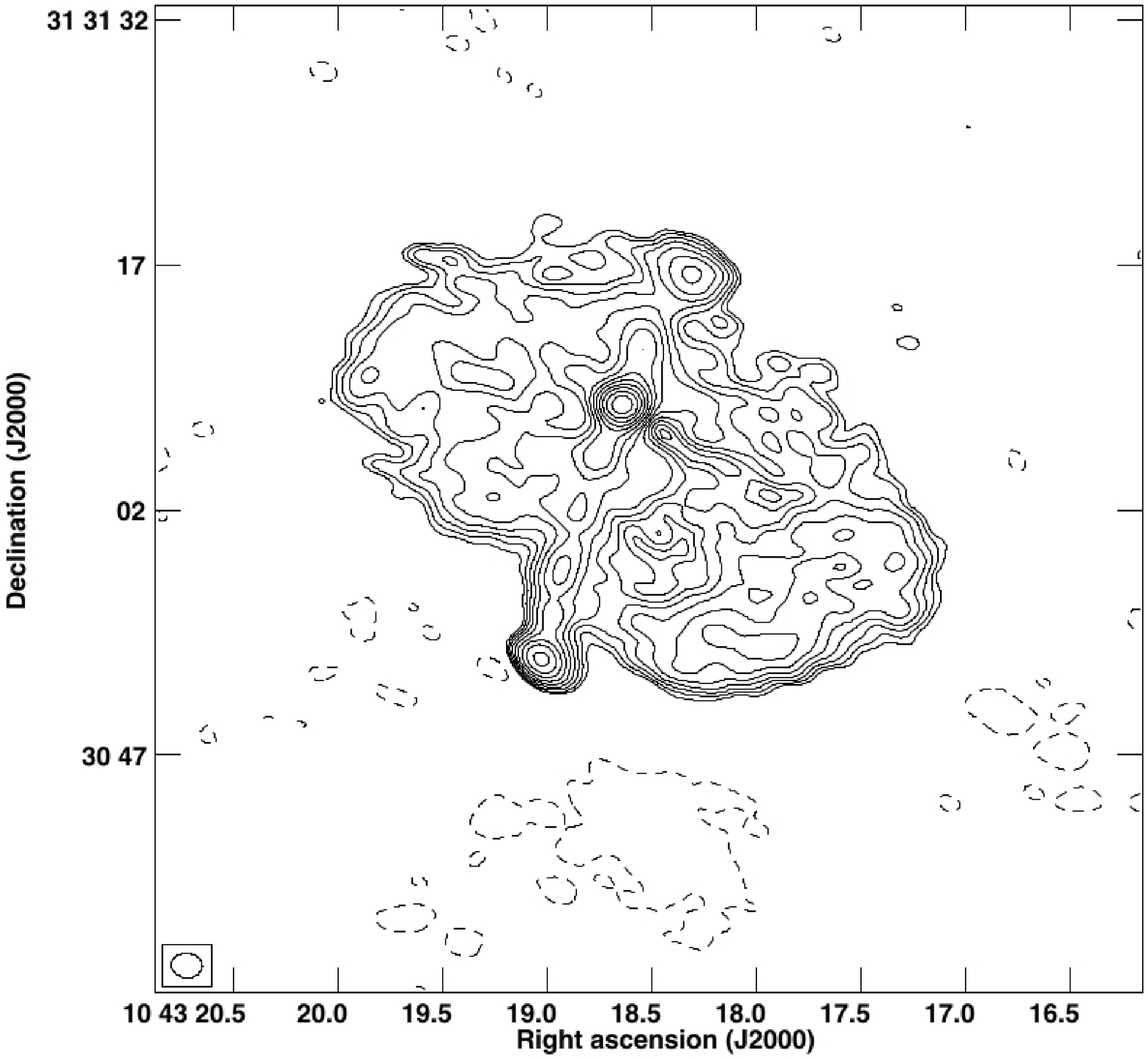} 
\caption{610 MHz radio map of the source 3C 136.1 observed at GMRT \citep{Lal2007} 
(left panel); 1.4 GHz radio map of the source J1434+5906 observed at VLA 
\citep{Roberts2015} (middle panel); 
1.4 GHz radio map of the source J1043+3131 observed at VLA \citep{Roberts2015} (right
panel)
 }
   \label{fig:fig13}
\end{figure*}

We can compare the synthetic maps of the surface brightness distribution in
Fig. \ref{fig:fig12} with radio maps of X-shaped sources.
The synthetic map in the top-left
panel of Fig. 12, viewing angle of $90^\circ$ is reminiscent, for instance, of
the radio map 3C 136.1 (Fig. \ref{fig:fig13}, left panel, from
\citealt{Lal2007}). Considering the middle-left panel of Fig. 12, viewing angle
of $45^\circ$, the synthetic map can be compared to the morphology displayed
by the source
J1434+5906 (from \citealt{Roberts2015}) and shown in Fig. \ref{fig:fig13}
(middle panel). The bottom-left panel of Fig. 12, viewing angle of $20^\circ$,
compares nicely with the source J1043+3131 (again from \cite{Roberts2015}) and
shown in Fig. \ref{fig:fig13} (right panel).
Within our model, the variety of morphologies displayed by X-shaped sources
can the be interpreted, beside the general effects of the the main physical
parameters, by the combination of different viewing angles and the
orientation of the jet direction with respect to the major axis of the host
galaxy. In particular,  sources with a morphology
similar to  J1043+3131  are not usually classified as classical X-shaped, however our 3D results
show that X-shaped sources can indeed take this morphology if the line of sight form a small angle with the jet direction.

\begin{figure*}[htbp]
   \centering
   \includegraphics[width=6cm]{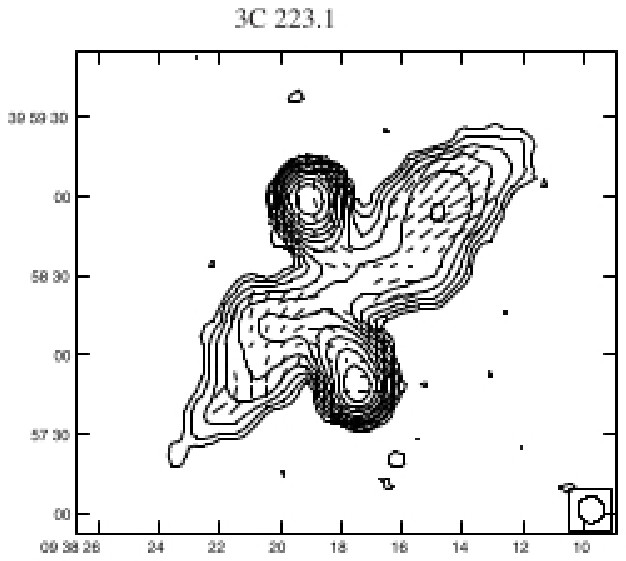} 
   \includegraphics[width=6cm]{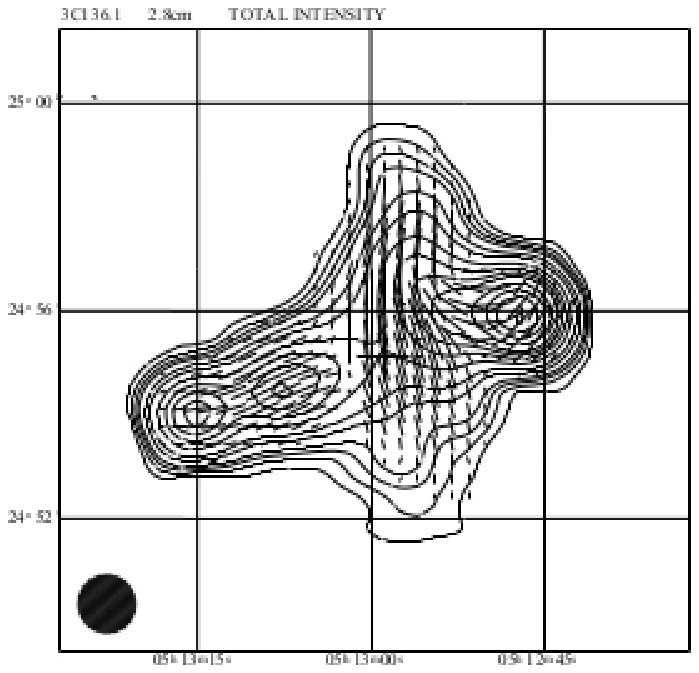} 
\caption{Polarization maps of the source  3C 223.1 \citep{Dennett-Thorpe2002} (left panel)
and of 3C 136.1 (\citet{Rottmann2001}) (right panel)
 }
   \label{fig:fig14}
\end{figure*}

From our polarization maps we see that the magnetic field is directed along
the jets and along the wings, becoming transverse only at the edges. This kind
of morphology can be observed also in actual sources, as we can see in Fig.
\ref{fig:fig14}, where we show the polarization map of 3C 223.1 (from
\cite{Dennett-Thorpe2002}, left panel) and the one of 3C 136.1 (from
\citet{Rottmann2001}, right panel). These maps have to be compared with the
right panels of Fig. \ref{fig:fig12}.  

\begin{figure*}[htbp]
   \centering
\includegraphics[width=15cm]{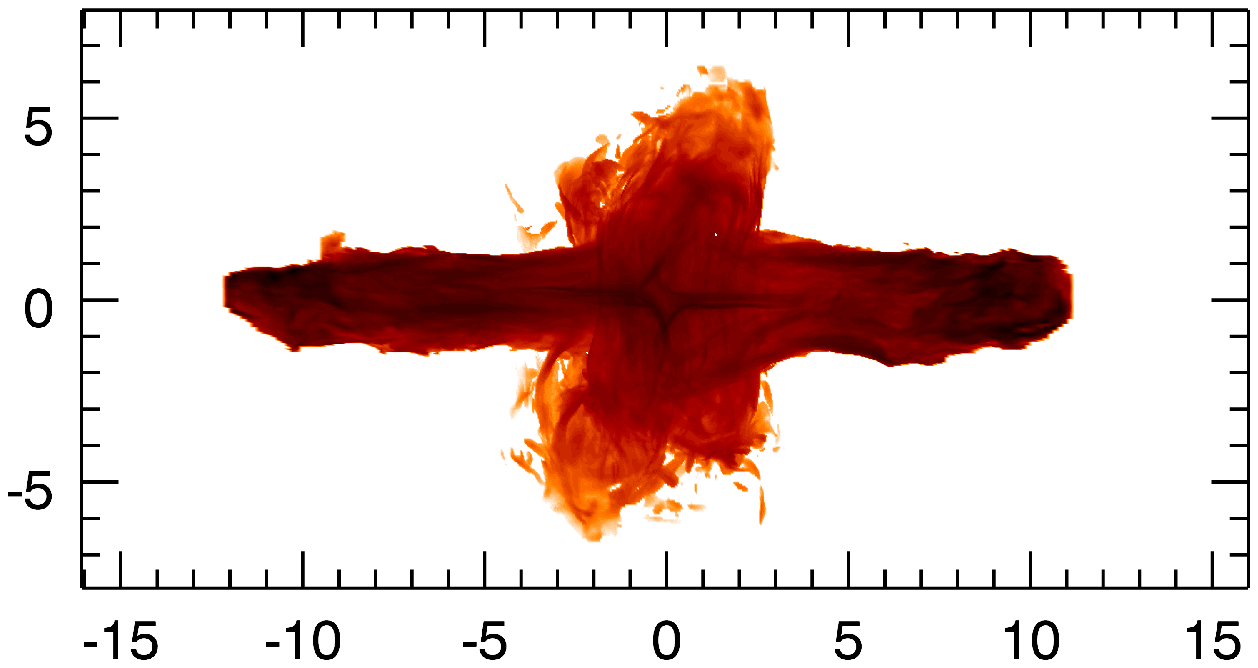} 

\caption{ Synthetic radio surface brightness  map, for case F, computed assuming local equipartition between the magnetic 
energy and the energy in non-thermal particles.  The line of sight makes an angle of  
$90^\circ$  with the jet direction and the map corresponds to the top left panel in Fig. \ref{fig:fig12}.  The 
brightness map, as in  Fig. \ref{fig:fig12} has a logarithmic scaling and cover a range of four order of magnitude.
 }
   \label{fig:fig15}
\end{figure*}

The emission here is treated in a simplified way and, since we do not model the 
evolution of the emitting non-thermal particles, we have to give a prescription for
the their density and distribution. Different prescriptions may lead to some differences in the 
synthetic maps.  We have then considered  an  alternative in which we prescribe local equipartition 
between the magnetic energy and the energy of non-thermal particles. In Fig. \ref{fig:fig15} we show, for this choice, 
the synthetic map of the surface brightness distribution for line of sight at $90^\circ$ with the jet direction, corresponding to
the top left panel in Fig.\ref{fig:fig12}. Comparing the two maps, we can observe an overall similarity, 
the main differences are in the jets and in the central part of the wings which  show a somewhat higher brightness.

A final remark concerns the distribution of the offset between the host major
axis and the wings. \citet{Capetti2002} and \citet{Gillone16} found that in
only one source (out of 31) the wings form an angle smaller than 40$^o$ and
this is the key observing result that motivated the simulations presented in
this paper. However, this finding carries further information on the wings
structure. In fact, in case the wings were linear and essentially one
dimensional features, a significant fraction of X-shaped sources should show,
due to projection effects, small misalignments; this would occur when the
plane defined by the wings axis and the major axis of the host forms a large
angle with the plane of the sky. Our simulations show instead that the wings
form a flattened emitting region aligned with a plane including the host minor
axis; the projection of such essentially two-dimensional structure leads to a
narrower range of offsets, as observed.

\section{Summary}
%
%

We presented 3D relativistic magnetohydrodynamic simulations of bidirectional
jets propagating in a triaxial density distribution in order to investigate
the mechanism of formation of X-shaped radiogalaxies. In Paper I we proposed a
hydrodynamic mechanism for the formation of this class of sources and we
performed 2D simulations of the propagation of a jet in an ellipsoidal density
distribution. We showed that, when the jet propagates along the major axis of
the distribution a winged structure can form. Here we extended our previous
results, relaxing the axial symmetry, introducing magnetic field and
relativistic effects. We investigated, in particular, under which conditions
of jet power, Lorentz gamma factor, and density ratio, an X-shaped structure
arises. We showed that jets with a power larger than $10^{45}$ erg s$^{-1}$
have to be non relativistic in order to form a winged structure, while
relativistic jets can form a winged structure only if they have $L_j
\lesssim 10^{44}$ erg s$^{-1}$. By applying a scaling relation between jet and
radio power, we estimated that the maximum radio luminosity of a winged source
is $\sim 3 \times 10^{43}$ erg s$^{-1}$, a value compatible with the
observations of X-shaped radio sources.  Moreover we considered a case in
which the jet is misaligned with respect to the major axis of the density
distribution and showed that an angle of $30^o$ is still favourable to the
formation of an X-shaped structure.

\citet{Hodges-Kluck2011} also performed three-dimensional simulations for a
configuration similar to ours, they did not include magnetic field and
relativistic effects.  An important difference between their analysis and ours
is the range of spatial and temporal scales considered. We focus on the first
phases of the evolution limiting to time scales of the order of $10^6$ yrs and
spatial scales of the order of $10$ kpc, when the jet still propagates within
the galactic atmospheres, while they consider the full life of the radiosource
and study its evolution when the jet is switched-off, the spatial scales are
also larger, and the jet propagates in the cluster environment. In this sense
the two approaches can be considered as complementary: the first phases of the
evolution, that we consider, are essential for the formation of an X-shaped
radiosource and the conditions that we find are necessary conditions, while
the results obtained by \citet{Hodges-Kluck2011} give important insights on the
subsequent long term evolution.

We have also calculated the brightness distribution producing synthetic
emission maps. Being the simulated sources intrinsically three-dimensional,
i.e. displaying a triaxial structure, the same objects may appear differently
as seen from varying viewing angles. We show that different observed
morphologies of X-shaped sources can be the result of similar structures
viewed under different perspectives. By considering both the structure of the
wings and the triaxiality of their hosts, the relative orientation of radio and
optical structures has a complex dependence on the projection; this leads not
only to favour large offsets between the wings and the host major axis, but
also to the observed paucity of offset smaller than $\sim 40^o$.

Finally we remark that we  carried out a simplified treatment of the emission. In reality, one should consider the evolution of a distribution of relativistic electrons, carried along the jets and streaming in the wings, that lose energy via synchrotron losses and adiabatic expansion, and gain energy by shock acceleration and/or via turbulent acceleration by IInd order stochastic Fermi-like process \citep[see][]{Vaidya17}. Our results  show that shocks in the wings are quite weak and therefore they will
most likely contribute in a limited way to particle reacceleration, however reacceleration by turbulence may be more effective.  Further studies are then required to examine the behavior of the spectral index in the wing, compared to observations.

\section*{Acknowledgments}
We acknowledge the CINECA award under the ISCRA initiative, for the availability of high performance computing resources and support.  This work has been partly funded by a PRIN-INAF 2014 grant.

\bibliographystyle{aa}
\bibliography{X-shape}

\appendix
\section{Surface brightness and polarization maps}
\label{ap:maps}

 The surface brightness radio maps and polarization maps  have been obtained by integrating along the line of sight respectively the synchrotron emissivity $j$ and the Stokes parameters $Q$ and $U$. The synchrotron emissivity in the comoving frame, for a power law distribution of relativistic electrons is \citep{DelZanna06}
\begin{equation}
j' (\nu') = K | \vec{B}'  \times \vec{n}' |^{\alpha+1} \nu'^{-\alpha} 
\end{equation}
where all primed quantities are computed in the proper frame, $\vec{n}'$ represents the versor  corresponding to the line of sight direction, $K$ is a constant proportional  to the number density of relativistic electrons and $-(2 \alpha + 1)$ is the slope of the energy distribution of relativistic particles. In order to obtain the emissivity in the observer frame, we have to take into account relativistic corrections and we get
\begin{equation}
j(\nu) = K | \vec{B}'  \times \vec{n}' |^{\alpha+1}  D^{\alpha + 2} \nu^{-\alpha} 
\end{equation}
where $D$ is the Doppler factor
\begin{equation}
D = \frac{1}{\gamma(1- \boldsymbol \beta \cdot \vec{n})}
\end{equation}
Considering a cartesian reference frame in which $X$ is along the line of sight and $Y$ and $Z$ are in the plane of the sky, the surface brightness can be obtained as
\begin{equation}
I(\nu,Y,Z) = \int^{\infty}_{-\infty} j(\nu, X, Y, Z) dX
\end{equation}
 and the Stokes parameters $Q$ and $U$ are given by
 \begin{equation}
Q(\nu,Y,Z) =\frac{\alpha + 1}{\alpha + 5/3} \int^{\infty}_{-\infty} j(\nu, X, Y, Z) \cos 2 \chi dX
\end{equation}
 \begin{equation}
U(\nu,Y,Z) = \frac{\alpha + 1}{\alpha + 5/3}  \int^{\infty}_{-\infty} j(\nu, X, Y, Z) \sin 2 \chi dX
\end{equation}
 where $\chi$ is the local polarization position angle, measured clockwise from the $Z$ axis. From $Q$ and $U$ we can then get the observed polarization direction.  In our calculations we assumed the density of relativistic electrons to be proportional to the local fluid density and the spectral index $\alpha = 0.5$.

\label{lastpage}
\end{document}